\documentclass[iop]{emulateapj}
\usepackage{apjfonts}
\usepackage{graphicx}
\usepackage{epstopdf}
\usepackage{ulem}
\usepackage[utf8]{inputenc}
\usepackage[T1]{fontenc}
\usepackage{url}
\usepackage{hyperref}
\usepackage{mdwlist}
\usepackage{amsmath} % or simply amstext
\usepackage{color}

\makeatletter
\def\url@leostyle{%
 \@ifundefined{selectfont}{\def\UrlFont{\sf}}{\def\UrlFont{\small\ttfamily}}}
\makeatother
\begin{document}

\newcommand{\ls}{{_<\atop^{\sim}}}
\newcommand{\gs}{{_>\atop^{\sim}}}
\def \spose#1{\hbox  to 0pt{#1\hss}}  
\def \ls{\mathrel{\spose{\lower 3pt\hbox{$\sim$}}\raise  2.0pt\hbox{$<$}}}
\def \gs{\mathrel{\spose{\lower  3pt\hbox{$\sim$}}\raise 2.0pt\hbox{$>$}}}
\newcommand{\Ha}{\hbox{{\rm H}$\alpha$}}
\newcommand{\Hb}{\hbox{{\rm H}$\beta$}}
\newcommand{\OIII}{\hbox{[{\rm O}\kern 0.1em{\sc iii}]}}
\newcommand{\NII}{\hbox{[{\rm N}\kern 0.1em{\sc ii}]}}
\newcommand{\angstrom}{\textup{\AA}}
\font\btt=rm-lmtk10
\definecolor{red} {cmyk}{0.   , 1.   , 1.   , 0.   }
\definecolor{blue} {rgb}{0,0,1}
\newcommand{\ledit}[1]{\textcolor{red}{\textbf{#1}}}

%% ------------------------------------------------------------------
%% TITLE
%% ------------------------------------------------------------------

\title{Galaxy Zoo: Observing Secular Evolution Through Bars\**}

\shorttitle{Observing Secular Evolution Through Bars}

\shortauthors{Cheung et al.}

%% ------------------------------------------------------------------
%% AUTHORS
%% ------------------------------------------------------------------

\author{Edmond Cheung\altaffilmark{1}\dag, E. Athanassoula\altaffilmark{2}, Karen L. Masters\altaffilmark{3} \altaffilmark{4}, Robert C. Nichol\altaffilmark{3} \altaffilmark{4},  A. Bosma\altaffilmark{2}, Eric F. Bell\altaffilmark{5}, S. M. Faber\altaffilmark{1} \altaffilmark{6}, David C. Koo\altaffilmark{1} \altaffilmark{6}, Chris Lintott\altaffilmark{7} \altaffilmark{8}, Thomas Melvin\altaffilmark{3}, Kevin Schawinski\altaffilmark{9}, Ramin A. Skibba\altaffilmark{10}, Kyle W. Willett\altaffilmark{11} }

\altaffiltext{1}{Department of Astronomy and Astrophysics, 1156 High Street, University of California, Santa Cruz, CA 95064}
\altaffiltext{2}{Aix Marseille Universit\'e, CNRS, LAM (Laboratoire d'Astrophysique
de Marseille) UMR 7326, 13388, Marseille, France}
\altaffiltext{3}{Institute of Cosmology \& Gravitation, University of Portsmouth, Dennis Sciama Building, Portsmouth, PO1 3FX, UK}
\altaffiltext{4}{SEPnet, South East Physics Network, ({\tt www.sepnet.ac.uk})} 
\altaffiltext{5}{Department of Astronomy, University of Michigan, 500 Church St., Ann Arbor, MI 48109, USA}
\altaffiltext{6}{UCO/Lick Observatory, Department of Astronomy and Astrophysics, University of California, 1156 High Street, Santa Cruz, CA 95064}
\altaffiltext{7}{Oxford Astrophysics, Department of Physics, University of Oxford, Denys Wilkinson Building, Keble Road, Oxford OX1 3RH}
\altaffiltext{8}{Astronomy Department, Adler Planetarium and Astronomy Museum, 1300 Lake Shore Drive, Chicago, IL 60605, USA}
\altaffiltext{9}{Institute for Astronomy, Department of Physics, ETH Zurich, Wolfgang-Pauli-Strasse 27, CH-8093 Zurich, Switzerland}
\altaffiltext{10}{Center for Astrophysics and Space Sciences, Department of Physics, 9500 Gilman Drive, University of California, San Diego, CA 92093, USA}
\altaffiltext{11}{School of Physics and Astronomy, University of Minnesota, USA}

\altaffiltext{\dag}{ec2250@gmail.com}
\altaffiltext{\**}{This publication has been made possible by the participation of more than 200 000 volunteers in the Galaxy Zoo project. Their contributions are individually acknowledged at \href{http://www.galaxyzoo.org/Volunteers.aspx}{http://www.galaxyzoo.org/Volunteers.aspx}.}

\slugcomment{Accepted by ApJ}

%% ------------------------------------------------------------------
%% ABSTRACT
%% ------------------------------------------------------------------

\begin{abstract}

In this paper, we use the Galaxy Zoo 2 dataset to study the behavior of bars in disk galaxies as a function of specific star formation rate (SSFR), and bulge prominence. Our sample consists of 13,295 disk galaxies, with an overall (strong) bar fraction of $23.6\pm 0.4\%$, of which 1,154 barred galaxies also have bar length measurements. These samples are the largest ever used to study the role of bars in galaxy evolution. We find that the likelihood of a galaxy hosting a bar is anti-correlated with SSFR, regardless of stellar mass or bulge prominence. We find that the trends of bar likelihood and bar length with bulge prominence are bimodal with SSFR. We interpret these observations using state-of-the-art simulations of bar evolution which include live halos and the effects of gas and star formation. We suggest our observed trends of bar likelihood with SSFR are driven by the gas fraction of the disks; a factor demonstrated to significantly retard both bar formation and evolution in models. We interpret the bimodal relationship between bulge prominence and bar properties as due to the complicated effects of classical bulges and central mass concentrations on bar evolution, and also to the growth of disky pseudobulges by bar evolution. These results represent empirical evidence for secular evolution driven by bars in disk galaxies. This work suggests that bars are not stagnant structures within disk galaxies, but are a critical evolutionary driver of their host galaxies in the local universe ($z<1$).

\end{abstract}

%% KEYWORDS
%%

%% ------------------------------------------------------------------
%% INTRODUCTION
%% ------------------------------------------------------------------

\section{Introduction} \label{sec:introduction}

Stellar bar-shaped structures within galaxies, or more simply `bars', have been known to exist since the days of Edwin Hubble. With only the 100 inch telescope at Mount Wilson, Hubble accurately surmised that bars were abundant in the local universe. So abundant, that he devoted a major part of his classification scheme, the Hubble sequence of galaxies \citep{hubble36}, to barred galaxies. Decades later, infrared and optical studies have confirmed that many galaxies have bars. Indeed, among local disk galaxies, as many as two thirds are barred \citep[e.g.,][]{mulchaey97, knapen00, eskridge00, kk04, menendez07, sheth08}.

\setcounter{footnote}{11}

Bars have an important influence on galaxy evolution. The presence of bars has been linked to the existence of spiral arms, rings \citep{sanders76, simkin80, schwarz81}, and/or disky pseudobulges\footnote{Bulges created through secular evolution have been called both ``pseudobulges'' and/or ``disky bulges". For completeness, we will use the term  ``disky pseudobulges" throughout to represent such bulges in galaxies.} \citep{kk04, athanassoula05}. Bars have also been associated with an increase in central star formation \citep{hawarden86, dressel88, giuricin94, huang96, martinet97, martin97, ho97, ellison11, oh12, wang12}, the flattening of galactic chemical abundance gradients \citep{villacostas92, zaritsky94, martinroy94, williams12}, and, perhaps, active galactic nuclei (AGN) activity \citep{noguchi88, shlosman89, mulchaey97, laine02, martini03, laurikainen04, jogee06, hao09, oh12}.

Given that bars have an important influence on galaxy evolution,
two natural questions are ``how do bars form and evolve?'' and  
``how do they affect their host galaxies?'' 
A review of the theoretical work on bars is given by \cite{athanassoula12}, so we will only summarize here the parts that are most relevant to this work \citep[see also][]{sellwood93, sellwood13}. 
Many past theoretical works have shown that bars can redistribute the angular
momentum of the baryons and dark matter of a galaxy \citep[e.g.,][]{sellwood80, debattista00, holleybockelmann05}. 
The angular momentum is emitted mainly by stars at (near-)resonance in the bar region and absorbed mainly by (near-)resonant material in the spheroid (i.e., the halo
and, whenever relevant, the bulge) and in the outer disk 
\citep[hereafter A03]{lyndenbell72, tremaine84, athanassoula03}. 

A03 showed that the redistribution
of angular momentum is not merely a side-effect of
bars, but is, instead, a process that is closely coupled to the evolution of bars.
Specifically, the exchange of angular
momentum from the inner disk to the outer disk and/or
spheroid (bulge/halo) is the main driver of bar evolution. The efficiency
of the angular momentum exchange is primarily dependent
upon the mass distribution and velocity dispersion
of the disk and spheroid.
More angular momentum can be redistributed if the spheroid mass density at the
location of the resonances is high, leading to stronger bars (A03).
The second factor governing 
the efficiency of angular momentum exchange is the velocity
dispersion. In lower velocity dispersion (lower temperature) disks
and spheroids, resonances can emit or absorb more 
angular momentum than in cases with high velocity dispersion,
thereby making the transfer of angular momentum more efficient (A03; \citealt{sheth12}).

This redistribution of angular momentum allows bars to drive 
gas, and to a lesser extent, stars, to the centers of galaxies
\citep{matsuda77, simkin80, athanassoula92, wada&habe92, wada&habe95,
  friedli93, heller94, knapen95, sakamoto99,sheth05}. This process is
responsible for the increase of bar length and strength and of the disk scale 
length (e.g., \citealt{hohl71}, , \citealt{debattista00}, A03, \citealt{oneil03}, \citealt{valenzuela03}, \citealt{debattista06}, \citealt{martinezvalpuesta06}, \citealt{minchev11}), the formation of a disky pseudobulge \citep{kk04, athanassoula05}, the increase of central star formation \citep{friedli93, martinet97, martin97}, and the dilution of abundance gradients \citep{friedli94, friedli95, martel13}. This process is known as secular evolution \citep{kk04, kormendy12}.

It has been shown that bar formation and evolution is also dependent
on the gas content in the galaxy \citep[e.g.,][]{sholsman93, berentzen98, berentzen07, villavargas10}. More recent simulations -- with a
multi-phase description of the gas, including star formation, feedback
and cooling, and a sufficiently large number of particles to describe
adequately the gas flow -- have shown that bars form later in
simulations with a larger gas fraction \citep[hereafter AMR13]{athanassoula13}.

Recent observational works have begun to test many of these predictions. For example, \cite{masters11} used classifications from Galaxy Zoo 2 (see \S\ref{data:gz}), to show that the fraction of disk galaxies that possess a bar (bar fraction) increases in redder disk galaxies \cite[see also][]{skibba12}. This result was confirmed by \cite{lee12}, who also used a large sample of galaxies from the Sloan Digital Sky Survey (SDSS), but with their own classifications (combining a mix of visual and automated methods). Assuming that galaxy color is closely related to galactic gas content \citep[e.g.,][]{catinella10, saintonge11}, then this is consistent with the expected effects of gas on bar formation and evolution. Indeed, using a sample of Galaxy Zoo 2 bars with HI measurements from the ALFALFA survey, \cite{masters12} found that bar fraction correlates strongly with HI content. In that sample, more bars were found in the gas poor disk galaxies, even at fixed color or stellar mass.

Alternatively, \cite{barazza08} and \cite{aguerri09} found different results using samples of SDSS galaxies with bars identified from ellipse fitting methods.  Both of these works found that bar fractions were larger for the bluer (and presumably more gas rich) galaxies in their samples. However, \cite{nair10a, nair10b} suggest a way to reconcile these results which came from samples of disk galaxies with very different selections; notably \citealt{barazza08} and \citealt{aguerri09} selected only blue galaxies as disks, and included lower redshift and less massive galaxies than were present in \citealt{masters11}, \citealt{masters12}, or \citealt{lee12}. The sample of \cite{nair10a}, which probed a wide range of stellar mass, suggested that bar fraction is bimodal with disk galaxy color -- having peaks both towards the bluer and redder disk galaxies$\footnote{\citealt{masters11} also commented on a possible upturn in bar fraction for the bluest galaxies in their sample}$. \cite{nair10a} suggest this trend may reveal two distinct types of bars, namely weak bars are predominantly found in lower mass and more gas rich (and bluer) spirals, while stronger bars are more common in massive, redder and gas poor disks.

In addition to the dependence on galaxy color, bar fraction has also been found to depend on inner galactic structure. \cite{masters11} found that bar fraction was correlated with fracDeV, which is a parameter measured by the SDSS representing the fraction of the best-fit light profile that originates from the de Vaucouleurs fit to the profile, as opposed to an exponential fit. \cite{lee12} also found that the bar fraction was highest at moderate central velocity dispersion. However, \cite{barazza08} found that barred galaxies are most likely to exist in galaxies with low S\'ersic indices while \cite{aguerri09} found that bars are most likely to exist in galaxies with low concentration indices. Although these results appear conflicting, they all show that the presence of a galactic bar is influencing the inner structure of these galaxies.

While the trends of bar fraction can reveal aspects of bar formation and evolution, bar fraction is crude as it hides information on the bar itself. According to A03, the characteristics of a bar (e.g., long or short) can be used as tracers of bar evolution. Therefore, a common bar property that has been studied in the literature is bar length. \cite{athanassoulaM02} and A03 predicted that the presence of a bulge will result in a longer and more evolved bar. Comparing this
prediction to previous observational works shows a good consensus;
early-type disk galaxies do indeed have longer bars
\citep{kormendy79, athanassoula80, martin95, elmegreen85,
regan97}. Larger samples and/or infra-red imaging
continues this agreement \citep{laurikainen02, erwin05, laurikainen07, menendez07,
elmegreen07,  aguerri09, gadotti11, hoyle11}.

In this paper, we use the Galaxy Zoo 2 dataset \citep{masters11, hoyle11, willett13} to investigate how the likelihood of a galaxy hosting a galactic bar depends on two important factors, namely the gas content of the galaxy and its inner galactic structure. 
We perform the same investigation with bar length and compare both of these sets of relationships to theoretical predictions, which will not only give us a better understanding of bar formation and evolution, but also a better understanding of how bars affect their host galaxy.

We begin in \S2 by describing all the data used in the paper, while the main observational results are presented in \S3. We compare our results with several theoretical simulations in \S4 and discuss our work, and these comparisons, in \S5. We conclude in \S6. In Appendix A, we discuss the completeness of our sample. We assume a cosmological model with $H_{0} = 70$ km s$^{-1}$ Mpc$^{-1}$, $\Omega_{m} = 0.30$ and  $\Omega_{\Lambda} =0.70$ throughout this paper.  

\begin{deluxetable}{lcc}
\tabletypesize{\scriptsize}
\tablewidth{3in} 
\tablecaption{Sample Selection}
\tablehead{   
\colhead{Criterion} &
\colhead{GZ2D} &
\colhead{BL} \\
\colhead{} &
\colhead{\#} &
\colhead{\#}
}
\startdata
 Galaxy Zoo 2* & 295,305 & 3,150 \\  
 $0.01<z<0.06$** &  76,336  & 2,674 \\
 $M_r<-20.15$  &  43,266   & 2,177 \\ 
 $b/a>0.5$   &  	28,540 & 1,753 \\
 $\frac{1}{4}$ answers bar question & 14,353 & 1,753 \\
 $p_{mg}<0.4$ & 14,038 & 1,734 \\
 GIM2D models $<$ 1 \arcsec offset & 13,328 & 1,655 \\  
 Quality GIM2D disks  & 13,328  & 1,159  \\
 MPA cross-match & 13,295 & 1,154 
 \enddata
 \tablecomments{*See footnote 17. **We only consider galaxies with spectroscopic redshifts.}
\label{tab:sample_selection}
\end{deluxetable}

\section{Data} \label{sec:data}

This section lists all sources of data that this paper uses. In order to have a fully complementary dataset, we cross-matched every dataset, as described in each subsection, resulting in a successive reduction of the initial sample size. As a guide, our initial dataset is described in \S\ref{data:gz}, which derives from SDSS DR7 (summarized in \S\ref{data:sdss}). We list the sample totals at the end of each subsection, starting with \S\ref{data:gz}. Table \ref{tab:sample_selection} lists every major cut made to our two samples, and the resultant sample sizes.

\subsection{SDSS} \label{data:sdss}

All the galaxies used in our sample are drawn from the Main Galaxy Sample in the Legacy area of the SDSS Data Release Seven (SDSS DR7; \citealt{strauss02, abazajian09}). Where possible, we use the standard photometric and structural parameters provided by the SDSS pipeline. For example, we use the SDSS information to define a surface stellar mass density within a radius of one kiloparsec of the center of the galaxy, $\Sigma_{\rm1~kpc}^*$. We choose one kpc for this density as it matches the typical scale of bulges \citep{fisher10} and therefore, should be closely related to the bulges of most galaxies. 

In detail, $\Sigma_{\rm1~kpc}^*$ is created from the SDSS galaxy surface brightness profiles, {\btt profMean}, which is the mean surface brightness in a series of circular annulii, from the {\btt PhotoProfile} table in the {\btt CasJobs} website\footnote{\href{http://casjobs.sdss.org/CasJobs/}{http://casjobs.sdss.org/CasJobs/}}. In accordance with the SDSS recommendations\footnote{\href{http://www.sdss.org/dr7/algorithms/photometry.html}{http://www.sdss.org/dr7/algorithms/photometry.html}}, we take the inverse hyperbolic sine of each cumulative profile and fit them with a natural cubic spline. After transforming the spline fits back with a sine function, we differentiate the fits and obtain an estimate of the azimuthally averaged surface brightness profile. Finally, we compute the magnitude and color within one kpc for each galaxy from these profiles and convert them into a stellar mass through a color-dependent mass-to-light ($M_*/L_g$) ratio \citep[e.g.,][]{bell01}. Our $M_*/L_g$ relationship is derived from a linear fit to the rest-frame $g-r$ color from GIM2D (see \S\ref{data:gim2d}) and the $M_*/L_g$, where the stellar masses are taken from the MPA-JHU catalog (see \S\ref{data:mpa}) and the $g$-band luminosity is taken from GIM2D models.

One concern is that the one kpc radius aperture is smaller than the typical seeing of SDSS. However, an analysis of angular sizes of galaxies in our sample, which lies within the redshift range $0.01<z<0.06$ (see \S\ref{data:gz}), shows they are typically larger than the full width at half maximum (FWHM) of the SDSS point--spread function ($\sim 1.3\arcsec$ in the $r$-band; \citealt{abazajian09}). 

\begin{figure*}[t!] 
\centering
\includegraphics[scale=.82]{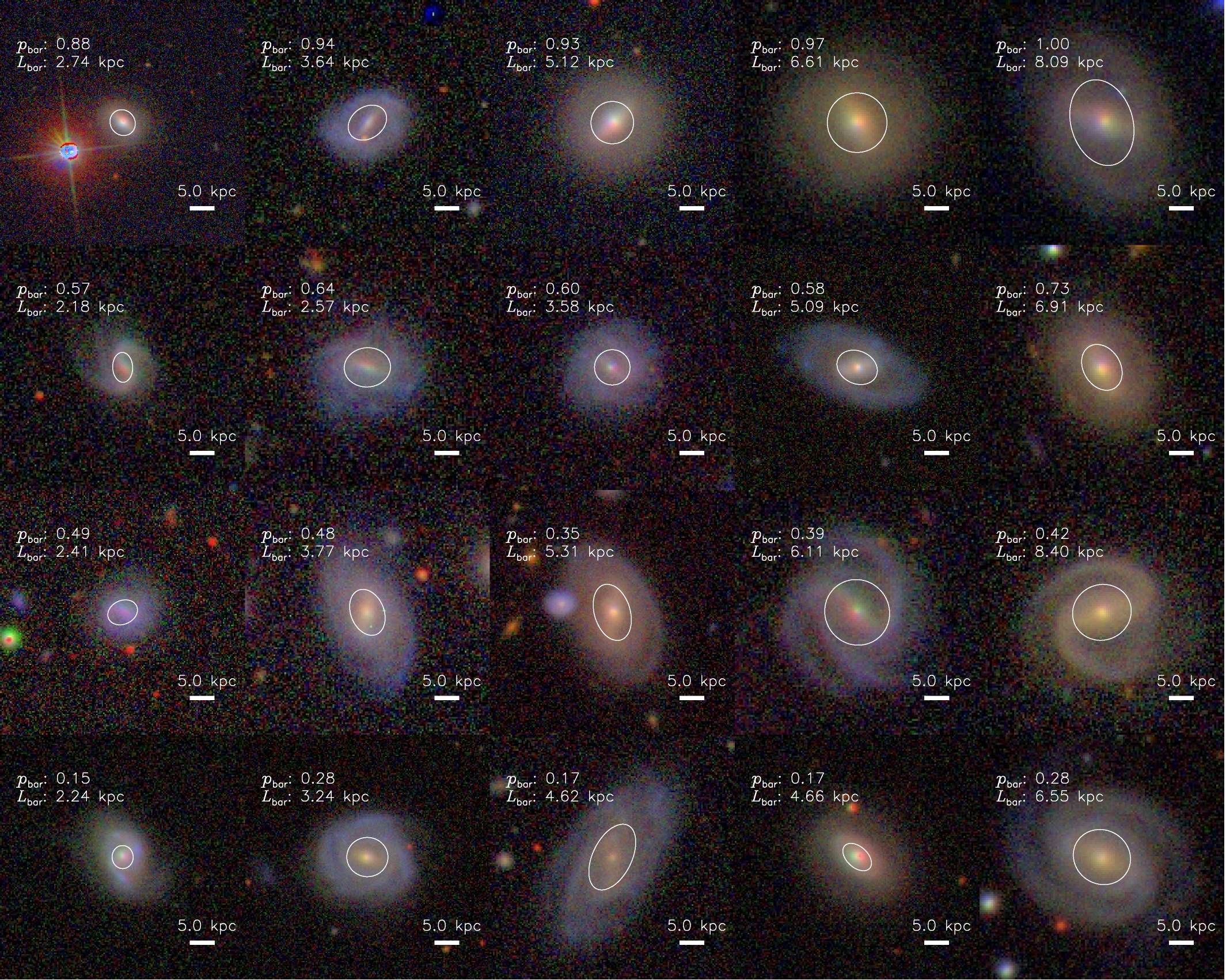}
\caption{A gallery of galaxies with a range of bar likelihood, $p_{\rm bar}$, and bar length, $L_{\rm bar}$. Each row is ordered by increasing bar length. The ellipse drawn over each galaxy represents the GIM2D disk model at $r_{\rm d}$. Visually, the bars generally extend out the disk scale length, consistent with \cite{combes93}. The physical scales of every image are the same ($\pm$ 1 pixel). 
\label{fig:bargallery}}
\end{figure*} 

\subsection{Galaxy Zoo} \label{data:gz}

Galaxy Zoo is a citizen science project that enlisted hundreds of thousands of volunteer ``citizen scientists" to make morphological classifications of nearly one million galaxies \citep{lintott08,lintott11}. The initial Galaxy Zoo project asked the public to classify galaxies as elliptical, spiral, or merger. With the Galaxy Zoo 2 project \citep{willett13}, the citizen scientists were asked to make more detailed classifications of approximately 304,000 galaxies.

The final product of Galaxy Zoo 2 is a table of morphological likelihoods, including the likelihood that a bar is present in each galaxy, as represented by $p_{\rm bar}$, e.g., if 5 out of 10 scientists classified a galaxy as having a bar, the galaxy would be assigned a bar probability of $p_{\rm bar}=0.5$. These raw probabilities are then adjusted to account for the reliability of each user through an iterative weighting scheme that ``down-weights" classifications from unreliable users (typically a few percent of the population). We also apply a correction to the likelihoods to account for the deterioration of the image quality due to increasing distance of galaxies, i.e., we assume galaxies of a similar luminosity and size will share the same average mix of morphologies regardless of redshifts. This also assumes there is no significant evolution within the SDSS at these low redshifts, which is probably reasonable \citep{bamford09, willett13}. Therefore, throughout the paper, we will only use these corrected, or ``debiased", bar likelihoods and will calling them $p_{\rm bar}$ for convenience. 

In \citet{masters11} and \citet{masters12}, barred galaxies were selected using $p_{\rm bar}\ge0.5$. This threshold delivered a high purity of barred galaxies in comparison with other barred galaxies sample, e.g., almost all galaxies with $p_{\rm bar}\ge0.5$ were classified as possessing a strong bar by \cite{nair10a} \citep[see Appendix A of][]{masters12}. Weaker bars in \cite{nair10a} were found to correspond to $0.3\le p_{\rm bar}\le0.5$ \citep{masters12, willett13}.

In this work, we choose to use $p_{\rm bar}$ as a bar likelihood, rather than as a bar threshold. This method has been used before with Galaxy Zoo classifications \citep[e.g.,][]{bamford09, skibba09, skibba12}. Our results are in qualitative agreement with other GZ results who used bar fractions, e.g., if we adopt a bar threshold of $p_{\rm bar}=0.5$, we find an overall bar fraction of  $23.6 \pm 0.4 \%$, which is similar to \cite{masters11}\footnote{The difference between our bar fraction and that of \cite{masters11} is due to the use of the weighted and debiased bar fractions from \cite{willett13} which were unavailable at the time of \cite{masters11}.}. 

Our initial sample is the Galaxy Zoo 2 dataset\footnote{This Galaxy Zoo 2 sample is comprised of the `original',  `extra', and `stripe82' sample in Table 1 of \cite{willett13}. These data are available at \href{http://data.galaxyzoo.org}{http://data.galaxyzoo.org}}. Following \cite{masters11}, we only select galaxies with spectroscopic redshifts in the range of $0.01<z<0.06$. In order to have a volume-limited sample, we only include galaxies with $M_{r}<-20.15$, where $M_{r}$ is the rest-frame absolute Petrosian $r$-band magnitude. This limit corresponds to the Galaxy Zoo 2 completeness Petrosian magnitude of 17 in the $r$-band \citep{willett13} at $z=0.06$. To ensure that our sample contains relatively face-on galaxies, we applied an axis ratio requirement of $b/a>0.5$ (this corresponds to inclination angles less than $\approx$ 60 degrees), where $b/a$ is the axis ratio from the GIM2D single S\'ersic model fit (see \S\ref{data:gim2d}). This requirement minimizes projection effects and thus results in more reliable bar classifications. This sample also requires that all galaxies have a {\btt Petro90} radius of $>3 \arcsec$. We have tested our results with a larger minimum radius requirement and find that our results are unchanged\footnote{We find that our results are unchanged when we restrict our sample to galaxies with global half-light radii (as measured by GIM2D) larger than $5\arcsec$.}. 

We also require that for each galaxy, at least a quarter of all its classifications involved answering the bar question,  {\it `Is there a sign of a bar feature...'} \citep{masters11}. In order to reach the bar question, however, a user must first classify a galaxy as a non-edge-on galaxy with a disk or some sort of feature (e.g., spiral arms, rings, bars). Assuming that most identified features are associated with a disk, then this last selection effectively ensures we have non-edge-on disk galaxies. 

Finally, we discard all merging galaxies from the sample since we are only concerned with isolated galaxies that have reliable photometric and structural measurements. According to \cite{darg10}, the Galaxy Zoo merging parameter, $p_{\rm mg}$, can identify merging galaxies with a cut of $p_{\rm mg}>0.4$; we adopt this threshold to eliminate merging galaxies. There is a total of 14,038 galaxies in the resulting sample, which we will call the Galaxy Zoo 2 Disk (GZ2D) sample.

\begin{figure*}[t!] 
\centering
\includegraphics[scale=.82]{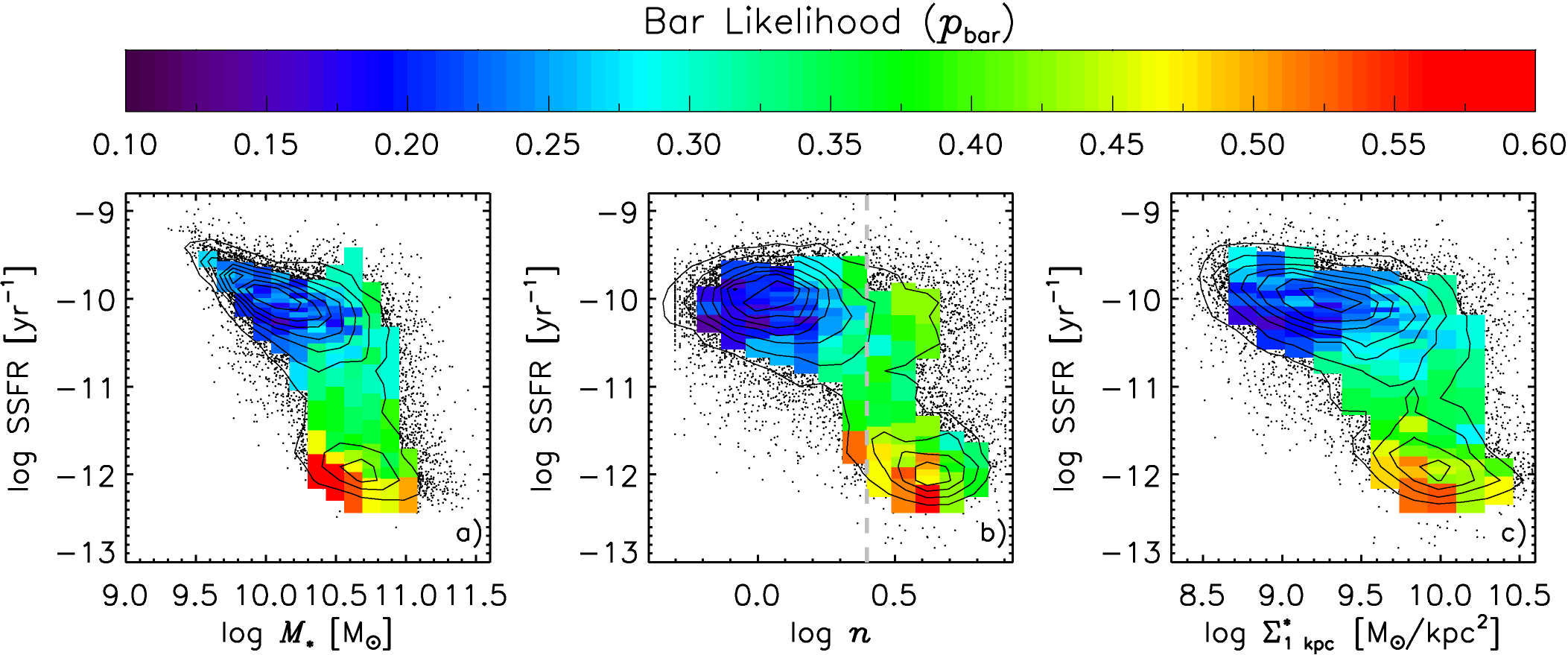}
\caption{Average bar likelihood, $p_{\rm bar}$, in bins of: {\it a)} SSFR vs. stellar mass $M_*$, {\it b)} SSFR vs. S\'ersic index $n$, {\it c)} SSFR vs. central surface stellar mass density $\Sigma_{\rm 1~kpc}^*$. Each bin is adjusted so that it contains $\sim100$ galaxies (individual points are shown for poorly populated bins). Bin colors indicate the average value of $p_{\rm bar}$ in the bin (see color bar at top), while the contours show the density of points. The gray dashed vertical line in panel b represents the division between galaxies containing disky pseudobulges ($n<2.5$) and classical bulges ($n>2.5$; see \citealt{drory07}). This plot shows that the trends of bar likelihood with galaxy properties depend on the SSFR of the galaxies. The relationship of $p_{\rm bar}$ with $n$ and $\Sigma_{\rm 1~kpc}^*$ is bimodal with SSFR. 
\label{fig:barfrac}}
\end{figure*}

We carefully review here the make-up of our sample to avoid confusion with comparisons with other disk, spiral or late-type selections based on GZ morphologies. The disk galaxy selection presented herein possibly includes a fraction of very early-type disks galaxies (Sa or S0) which would normally be included in a majority of early-type samples selected either by color, or central concentration. This results in our diverse disk galaxy sample showing bimodality in their optical color-mass diagram (Fig.~\ref{fig:barfrac}a). However, other Galaxy Zoo samples, that are more focused on late-type disks or spirals sample  (Sb, Sc or later) can be constructed using the GZ1 ``clean'' spiral criterion as first discussed in \cite{land08}, and most recently used in Schawinski et al. (2013, in preparation), but also through stricter limits in GZ2/GZ Hubble data. This more conservative late-type sample will be more dominated by ``blue cloud'' spirals and thus show less bimodality of their galaxy properties. 

In addition to this sample, we use a Galaxy Zoo 2 subsample that possesses additional bar length measurements. The bar lengths were visually measured by citizen scientists using a Google Maps interface described by \cite{hoyle11}. The bar lengths represent the lengths from one end to the bars to the other. In order to be consistent with previous works, who define it as the semi-major axis of maximum ellipticity in the bar region \citep[e.g.,][]{erwin05}, we will take half of the Galaxy Zoo 2 bar lengths and denote it $L_{\rm bar}$. This catalog requires at least 3 independent bar length measurements per galaxy; the mean of these independent bar length measurements gives $L_{\rm bar}$ of each galaxy. The vast majority of galaxies that were selected for this sample have $p_{\rm bar}\ge0.6$, i.e., this sample contains mainly strong bars \citep{masters12, willett13}. Of the GZ2D sample, there are 1,734 galaxies that have bar length measurements, which will now be referred to as the Bar Length (BL) sample.

We present a gallery of barred galaxies with a range of $p_{\rm bar}$ and $L_{\rm bar}$ in Fig.~\ref{fig:bargallery}. Each row is ordered by absolute bar length. 

\subsection{GIM2D} \label{data:gim2d}

Two-dimensional bulge+disk decompositions in the $g$ and $r$ bandpasses of over a million SDSS galaxies were performed with GIM2D by \cite{simard11}. Improvements to the sky background determinations and object deblending over the standard SDSS procedures led to more robust galactic structural parameters than those offered by the standard SDSS pipeline. 

Three different models were used in these decompositions: a pure S\'ersic model, an $n=4$ bulge + exponential disk model, and a S\'ersic (free-floating $n$) bulge + exponential disk model. The most important GIM2D parameter for the GZ2D sample is the galaxy S\'ersic index, $n$, from the pure S\'ersic model, i.e., the best-fitting single S\'ersic index for a given galaxy. The S\'ersic index has often been used to separate early-type and late-type galaxies and is widely regarded as a good proxy for bulge dominance \citep{blanton03, shen03, bell04, schiminovich07, drory07, bell08, wuyts11, bell12, wake12, cheung12}. 

\begin{figure}[t!]  
\centering
\includegraphics[scale=.7]{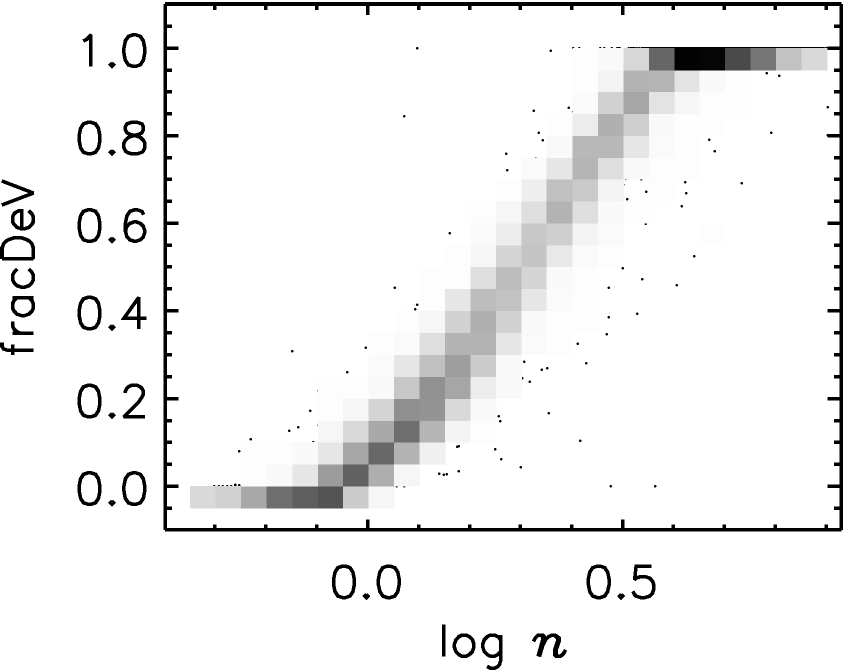}
\caption{Comparison of $n$ from the GIM2D single S\'ersic model fit and fracDeV ($r$ band) from the SDSS database. The cluster of galaxies at fracDeV=1 and fracDeV=0 accounts for $\sim27\%$ of the total GZ2D sample, indicating that there a loss of galaxy structural information in the fracDeV parameter.} 
\label{fig:nvsfracDeV}
\end{figure}

Although a similar parameter, fracDeV from the SDSS database, has been explored by previous works \citep[e.g.,][]{masters11, skibba12}, $n$ is a more common parameter in the literature and has been thoroughly studied \citep[e.g.,][]{graham05}. It is also the basis of most galaxy fitting programs (e.g., GALFIT, BUDDA, and GIM2D; \citealt{peng02, desouza04, simard02}), which allows for easier and more consistent comparisons to other works. For reference, we compare $n$ and fracDeV in Fig.~\ref{fig:nvsfracDeV} for our GZ2D sample. Clearly, the two parameters are correlated. However, the overdensity of galaxies at fracDeV=1, which accounts for $\sim15\%$ of the GZ2D sample, indicates that there is a saturation of galaxy structural information in the fracDeV parameter. Indeed, for fracDeV=1, $\log~n$ ranges from 0.5 to 0.9, corresponding to $n\sim3-8$. A similar effect occurs at fracDeV=0, which accounts for another $\sim12\%$ of the GZ2D sample. Our use of S\'ersic index in this paper should be more sensitive than fracDeV to the complicated structures of galaxies.

Another similar parameter is the Petrosian concentration index from SDSS. This parameter has been shown by \cite{gadotti09} to be a better proxy for bulge fraction than the global S\'ersic index. We would like to note, however, that the global S\'ersic indices that \cite{gadotti09} used were from the New York University Value-Added Galaxy Catalog \citep{blanton05a, blanton05b}, which fitted one-dimensional profiles extracted from two-dimensional images using {\it circular} annuli. The GIM2D fits were done using {\it elliptical} annuli, and are two-dimensional fits. As noted by \cite{simard11}, this difference in methodology, i.e., using circular and elliptical apertures, results in a systematic offset between the NYU and GIM2D galaxy half-light radius and galaxy S\'ersic index. At the request of the referee, we tested our results using R90/R50 in Appendix~\ref{appendix:r90r50} -- we find no major impacts to our conclusions.

The most important GIM2D parameter for the BL sample is the semi-major axis exponential disk scale length, $r_{\rm d}$; this is needed to properly scale the bar length. The disk scale length is available in both the $n=4$ bulge + exponential disk model and the S\'ersic bulge + exponential disk model; we use the latter model\footnote{We find no change in our main conclusions if we use the $n=4$ bulge + exponential disk model.}. As is noted in \cite{simard11}, the quality of the GIM2D bulge+disk decompositions is highly dependent on the spatial resolution and signal-to-noise of the SDSS images. Therefore, it is important to ensure that we only allow model fits that are reliable. However, since we are only concerned with $r_{\rm d}$, picking out reliable decompositions is not difficult. From Simard (priv. comm.), galaxy models with $B/T\le0.5$ (the $B/T$ from the S\'ersic bulge + exponential disk model) accurately model the disk component, and hence we consider all these galaxies. This is understandable since these galaxies are disk-dominated and their corresponding GIM2D models will likely yield reliable disk measurements. For models with $B/T>0.5$, Simard (priv. comm.) recommends considering only galaxies with $P_{pS}<0.32$, where $P_{pS}$ represents the probability that a bulge+disk model is $not$ required compared to a pure S\'ersic model \citep{simard11}. Thus $B/T>0.5$ galaxies that have a high probability of requiring a bulge+disk model are also considered. 

To avoid the effects of the SDSS point-spread function on the GIM2D disk model, we only allow disk models with  $r_{d} > 2\arcsec$. Furthermore, we impose a strict face-on requirement such that all GIM2D model disks have inclination angles of less than 55 degrees. This corresponds to axis ratios greater than 0.6, a parameter space that has been shown by \cite{macarthur03} to produce no systematic variations on $r_{\rm d}$ when using 2D galaxy decompositions. Finally, we require that the fractional errors on $r_{\rm d}$ ($r_{\rm d, error}/r_{\rm d}$, where $r_{\rm d, error}$ is the formal error of $r_{\rm d}$ from GIM2D) be less than $2\%$. This number is approximately two standard deviations above the average $r_{\rm d, error}/r_{\rm d}$ of the BL sample. We choose this conservative cut in order to include only quality disk models. 

We note that, although we only model a bulge and disk for these disk galaxies, $\sim24\%$ of which are strongly barred, previous works have shown that, while bulge parameters may be affected by the presence of a bar, the disk scale length is not significantly affected \citep{erwin05, laurikainen05}. This reliability is evident in the fact that our results are not sensitive to the choice of GIM2D bulge+disk model, i.e., both $r_{\rm d}$ from the $n=4$ bulge+disk model and the S\'ersic bulge+disk model produce the same results. Furthermore, the GIM2D formal errors on $r_{\rm d}$ are not significantly different from strongly barred systems ($p_{\rm bar}>0.8$) and non-barred systems ($p_{\rm bar} < 0.05$).

We impose a final cut that eliminates all GIM2D models where the centers are offset from the input science images by more than one arscecond. Large offsets like these usually represent a bad fit, and, indeed, upon visual inspection, we find that almost all these cases contained bright point sources within the galaxy and/or diffraction spikes from nearby stars. Matching the GZ2D and BL samples to the GIM2D catalogs leave us with 13,328 and 1,159 disk galaxies, respectively. 

\subsection{MPA-JHU} \label{data:mpa}

Stellar masses and star formation rates are taken from the MPA-JHU DR7 release\footnote{\href{http://www.mpa-garching.mpg.de/SDSS/DR7/}{http://www.mpa-garching.mpg.de/SDSS/DR7/}}. Stellar mass ($M_*$) estimates are calculated using the Bayesian methodology and model grids described in \cite{kauffmann03}. The models are fit to the broadband $ugriz$ SDSS photometry, instead of the spectral indices from the 3$\arcsec$ fiber aperture. These estimates are corrected for nebular emission and a \cite{kroupa01} initial mass function is assumed. 

Star formation rates (SFR) are based on the technique presented in \cite{brinchmann04}. For their `Star-Forming' class, which consists of 39,141 galaxies, they estimate the SFR from model fits that cover a wide range of star formation histories of several emission lines from the SDSS fiber. For `Low S/N Star-Forming' class, which contains 29,115 galaxies, they convert the observed H$\alpha$ luminosity into a SFR. And for the `AGN', `Composite' and `Unclassifiable' classes, which contain a total of 66,986 galaxies, they use the D4000 value to estimate SFR/$M_*$ and SFR. Aperture corrections follow the method of \cite{salim07}, resulting in the SFR of the entire galaxy. The specific star formation rate (SSFR), a parameter that will be used throughout the paper, is defined to be the SFR divided by stellar mass; it was calculated by combining the SFR and $M_*$ likelihood distributions as outlined in Appendix A of \cite{brinchmann04}.

Matching the GZ2D and BL samples to the MPA-JHU catalog brings our final sample to 13,295 and 1,154, respectively. A detailed discussion of the completeness of the GZ2D and BL samples is presented in Appendix~\ref{appendix:completeness}. We find that while we are missing some low-mass quiescent disk galaxies, the effect is small and does not affect our results.

\begin{figure*}[t!]  
\centering
\includegraphics[scale=.8]{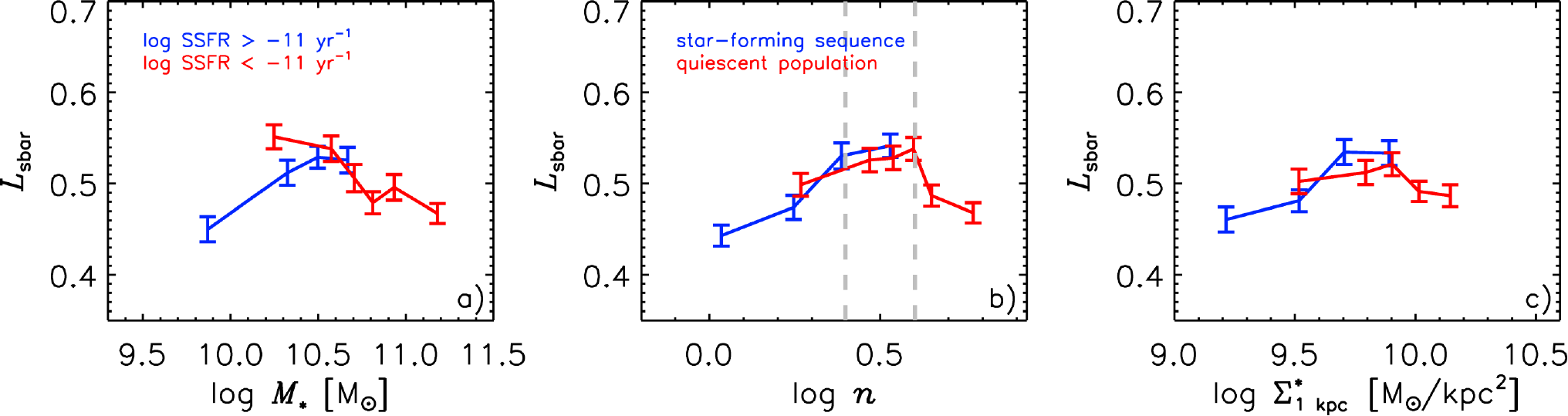}
\caption{Average $L_{\rm sbar}$ plotted against: {\it a)} $M_*$, {\it b)} $n$, and {\it c)} $\Sigma_{\rm 1~kpc}^*$. Galaxies were split by their star formation state, namely, log SSFR > -11 yr$^{-1}$ (star-forming; blue) and log SSFR <  -11 yr$^{-1}$ (quiescent; red). Each bin contains $\sim100$ galaxies. The error bars are given by $\sigma/\sqrt{N}$, where $\sigma$ is the standard deviation of $L_{\rm sbar}$ per bin, and $N$ is the total number of galaxies per bin. The vertical dashed lines in panel b are located at $\log~n=0.4$ ($n=2.5$) and $\log~n=0.6$ ($n=4$).
\label{fig:barlength_vs}}
\end{figure*}

\section{Results} \label{sec:results}

\subsection{Bar Likelihood Trends} \label{sub:barfrac}

In this section, we present the trends of bar likelihood with SSFR, stellar mass, and measures of bulge prominence. 

The three panels of Fig.~\ref{fig:barfrac} plot both galaxy density and average $p_{\rm bar}$ in a 2D plane of: SSFR versus stellar mass ($M_*$; panel a), SSFR versus global S\'ersic index ($n$; panel b), and SSFR versus central surface stellar mass density ($\Sigma_{\rm 1~kpc}^*$; panel c). The locations of the galaxies are shown by the contours. Bin sizes are adjusted so that they contain $\sim100$ galaxies each, and individual data points are shown for poorly populated bins. Each bin is colored by the average $p_{\rm bar}$ of the galaxies in it as indicated by the color bar. 

The well-known bimodality between galaxies (even for disk galaxies) in the star-forming sequence and those in the quiescent population is clear in our sample and affects not only the galaxies' SSFR, baryonic mass, and bulge properties, but also their likelihood of being barred \citep{masters11}. We find that there is a strong correlation between average $p_{\rm bar}$ and SSFR such that the average values of $p_{\rm bar}$ are larger for low SSFR disk galaxies (i.e., quiescent disk galaxies are more likely to host bars). The observed relationship between $p_{\rm bar}$ and SSFR is present even at fixed $M_*$, $n$, or $\Sigma_{\rm 1~kpc}^*$ (Fig.~\ref{fig:barfrac}), indicating that this relationship is nearly independent of these galaxy properties. 

Taking SSFR as a proxy for gas fraction \citep[e.g.,][]{kauffmann12} suggests that the underlying relationship is really between $p_{\rm bar}$ and gas content such that bar likelihood is increasing as gas fraction decreases. Similar trends between bar fraction and gas content were also observed by \cite{masters12}. 

We observe that the trends of the average bar likelihood with $M_*$, $n$, and $\Sigma_{\rm 1~kpc}^*$ depend on whether the disk galaxy is star-forming or quiescent, as is illustrated by Fig.~\ref{fig:barfrac}. Thus we look in more detail at the observed trends within the star-forming ($\log~ \rm SSFR > -11~ \rm yr^{-1}$) and quiescent ($\log~\rm SSFR < -11~ \rm yr^{-1}$) disk galaxy populations. We find: 
\begin{itemize} 
\item {\bf Stellar Mass, $M_*$ (Fig.~\ref{fig:barfrac}a)} -- There is a correlation between average $p_{\rm bar}$ and stellar mass within the star-forming disks such that $p_{\rm bar}$ is larger the larger their stellar mass. There is also an anti-correlation of $p_{\rm bar}$ with stellar mass within the quiescent population.  
\item {\bf S\'ersic Index, $n$ (Fig.~\ref{fig:barfrac}b)} -- For the star-forming sequence, $p_{\rm bar}$ is strongly correlated with $n$ (even more so than it is with $M_*$). Within the quiescent population, we see an inverse correlation between $p_{\rm bar}$ and $n$. This is an important point to note and might explain the contradictions between the results of previous studies, which found opposite trends of bar fraction with measures of bulge prominence from light profile shape (e.g., \citealt{masters11} compared to \citealt{barazza08}). Moreover, this observation is in good agreement with theoretical predictions of bar formation as will be described in \S\ref{sec:compare_theory}.
\item {\bf Central surface stellar mass density, $\Sigma_{\rm 1~kpc}^*$ (Fig.~\ref{fig:barfrac}c)} -- 
We find similar trends of $p_{\rm bar}$ with this parameter as between $p_{\rm bar}$ and $n$. Star-forming galaxies show a correlation between $p_{\rm bar}$ and $\Sigma_{\rm 1~kpc}^*$ (star-forming disks are more likely to host bars where the central density is higher), while quiescent galaxies show an anti-correlation (quiescent disks are more likely to host bars where the central density is lower)
\end{itemize}

\subsection{Bar Length Trends} \label{sub:barlength_results}

In this section we examine how bar length depends on galaxy properties. We define a scaled bar length, $L_{\rm sbar}$, as the bar length divided by a measure of disk size. We choose for this $2.2r_{\rm d}$ (2.2 semi-major axis exponential disk scale lengths) because this is where the rotation curve of a self-gravitating exponential disk reaches its maximum \citep{freeman70}. Hereafter, we will refer to the scaled bar length simply as the bar length unless stated otherwise.

Bars become longer over time as they transfer angular momentum from the bar to the outer disk and/or spheroid (halo and, whenever relevant, bulge). This secular evolution causes the host disk to expand and increase its scale length while the bar also grows. We will compare trends of bar length with those of $p_{\rm bar}$ to test if the trends we observe in the average value of $p_{\rm bar}$ in the galaxy population are due to the evolution of the bars, or the likelihood of bar formation in a galaxy.
 
Since the BL sample is more than an order of magnitude less numerous than the GZ2D sample, we find that breaking it up into small bins, as we did in Fig.~\ref{fig:barfrac} for $p_{\rm bar}$ results in no clear correlations. Since we found that the trends of $p_{\rm bar}$ had different properties depending on the SSFR of the galaxies, we split the BL sample into two subsamples (star-forming, or $\log ~\rm SSFR > -11~ \rm yr^{-1}$, and quiescent, or $\log ~\rm SSFR < -11~ \rm yr^{-1}$) to look at the trends of average $L_{\rm sbar}$. These trends are shown in Fig.~\ref{fig:barlength_vs}.  

This figure shows that in the star-forming sequence, the average value of $L_{\rm sbar}$ increases with all three properties ($M_*$, $n$, and $\Sigma_{\rm1~kpc}^*$). In the quiescent population we find that the average bar length decreases with $M_*$. Curiously we find that the average bar length increases with $n$ and $\Sigma_{\rm1~kpc}^*$ up to a maximum value at around $\log~n\approx0.6$ ($n\approx4$) and $\Sigma_{\rm1~kpc}^*\approx10^{9.8}$ M$_\odot$/kpc$^{2}$ respectively, where the trend reverses. 

\section{Comparison to Theory} \label{sec:compare_theory}

In this section, we compare our results in \S\ref{sec:results} with theoretical 
expectations of bar formation and evolution. We start with
a short summary of theoretical results.

\subsection{Theoretical Reminders} \label{sub:reminders}

One can distinguish (at least) two phases in the lifetime of a
bar: the formation phase and the secular evolution phase. 
AMR13 showed that these two phases are contiguous in gas-rich
cases, while, for gas-poor ones, they are not. In the latter case,
there are two further stages of relatively short duration in between the formation and
secular evolution phases\footnote{These two extra stages are related to
the bar buckling phase (i.e., the formation of a boxy/peanut
bulge), which is much less obvious in gas rich cases.}. 

\begin{figure}[t!] 
\centering
\includegraphics[scale=.7]{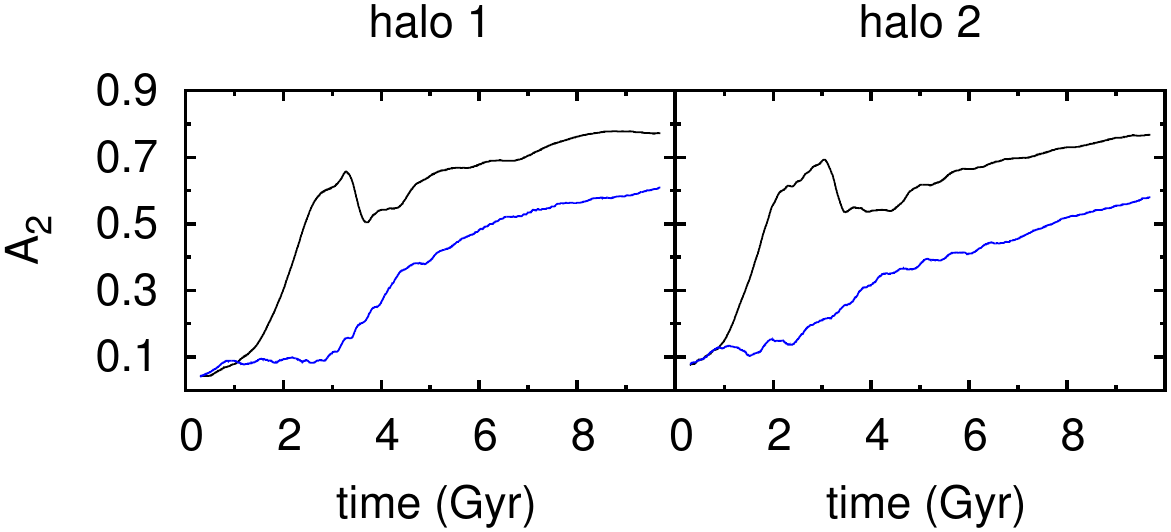}
\caption{Bar strength, A$_{2}$, which can considered as a rough proxy 
  for bar length,
  as a function of time for four simulations. 
  Two simulations have a gas-rich disk (blue lines) while the 
  other two have a gas-poor disk
  (black lines). The two panels correspond to different types of
  haloes: initially spherical (left panel) and initially triaxial
  (right panel). For a full description of these simulations and
  their results, see AMR13. These simulations show that 
  bars grow slower and are less strong in the gas-rich case.}
\label{fig:barevol}
\end{figure}

This is illustrated in Fig.~\ref{fig:barevol}, where we plot
the bar strength, A$_2$, which is closely related to bar length, 
as a function of time for four
simulations from AMR13. The two simulations in 
each panel have the same initial
mass and velocity distribution of the baryonic and dark matter
components. The only difference is the gas fraction, where
the black and blue lines represent gas-poor and gas-rich simulations,
respectively. The end of the bar formation phase is represented by the
time when the steep increase of A$_2$ terminates, which is at
times 2 -- 2.5 Gyrs for the 
gas-poor simulations, and around 4.5 Gyrs for the gas-rich ones. 
These simulations illustrate that gas slows down bar formation considerably 
(AMR13). This is due both to an increase in the duration 
of the pre-bar phase (i.e., the phase during which the disk can still
be considered as axisymmetric) and a 
decrease in the rate of the bar growth (i.e., an increase of the time it takes
for the bar to end its growth phase), both being compared to
the times of the equivalent phases in the gas-poor case. 

The secular evolution phase, however, starts roughly at 4.5
Gyrs for all cases. In general, the duration of these phases, 
as well as the increase of bar strength that they imply, depend on
the mass and velocity distribution of the baryonic and dark matter
components within the galaxy, as well as on the
gas fraction. Readers can find more information and a long list
of relevant references in a recent review by \cite{athanassoula12}.
It is also interesting to
note in Fig.~\ref{fig:barevol} that, for all times and in both phases, 
the bar in the gas-rich case is less strong than in the gas-poor one 
\citep[see also][]{berentzen07}.

Bar formation and evolution is influenced also by galactic bulges.
Bulges, however, are an inhomogeneous class of objects \citep{kormendy93,
kk04, athanassoula05}. Classical bulges 
have high S\'ersic indices, typically around
4, but certainly above 2. Disky pseudobulges, on the other hand, 
have low S\'ersic indices, typically around 1, and usually less than 2 \citep{fisher10}. 
The most popular scenario for the formation of disky
pseudobulges in barred galaxies is that they are due to stars, and
particularly, gas pushed inwards by the bar 
to the central parts of the disk. Here, the high density gas will
give rise to star formation, so that the disky pseudobulges should be
primarily composed of gas and young stars with a smaller fraction of
old stars. Their extent is typically of the order of 1
kpc \citep{athanassoula92, heller94, fisher10}\footnote{For completeness, we mention the boxy/peanut bulges, which, in fact, are part of the bar. Their S\'ersic
indices are smaller, or of the order of that of the disky pseudobulges. Given
that all our decompositions here include only one or two components
(\S\ref{data:gim2d}), and bars are not included, and our sample excludes highly inclined systems,
such bulges do not enter in our discussion. However, we do note that they may still be present in the sample and may not be well fit by our decompositions.}. 

\begin{figure*}[t!]  
\centering
\includegraphics[scale=.7, angle=90]{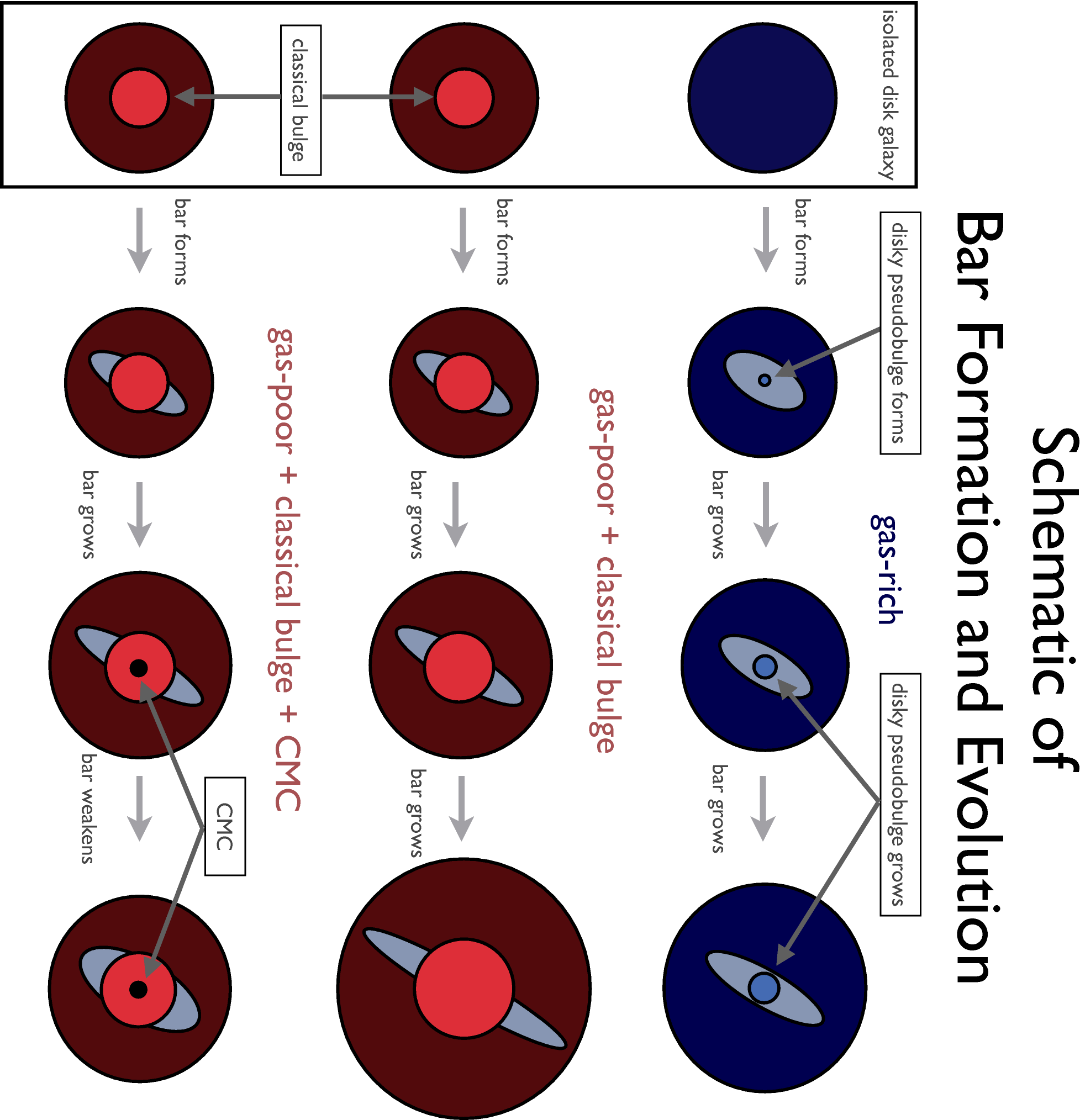}
\caption{A schematic diagram of bar formation and evolution. {\it Top row} illustrates the gas-rich scenario, in which a bar forms and grows over time. As the bar enters the secular evolution phase, a disky pseudobulge is created. The growth of the disky pseudobulge follows that of the bar. {\it Middle row} illustrates the gas-poor scenario with a classical bulge. The evolution of the bar in the gas-poor case is faster than that of the gas-rich case. {\it Bottom row} illustrates the gas-poor scenario with a classical bulge and a central mass concentration (CMC). The development of a CMC weakens the bar. 
\label{fig:bar_cartoon}}
\end{figure*}

These two different types of bulges have different dynamics and,
therefore, different effects on the bar formation 
and evolution phases. Classical bulges predate the bar, so they will
influence both phases. Their influence has many similarities to that
of the dark matter halo. Namely, they slow down bar formation in the first phase,
but, during the secular evolution phase, they help the bar grow by
absorbing angular momentum, leading to stronger bars (\citealt{athanassoulaM02}, A03). Thus, simulations predict that bars in galaxies with 
classical bulges should be stronger than bars in galaxies without classical bulges, 
assuming all other properties are the same.

On the other hand, disky pseudobulges in barred
galaxies are formed by material pushed inwards by the bar, i.e., they
do not predate the bar and thus cannot influence its formation
phase. Moreover, disky pseudobulges should not help the bar grow during
the secular evolution phase either, since they cannot absorb angular
momentum. This is because the radii of disky pseudobulges 
are considerably smaller than the corotation radius, and also because
disky pseudobulges are flat (spherical-like density distributions, like the
classical bulge or the halo, can absorb angular momentum). However,
although disky pseudobulges do not affect bar formation or evolution, bars
do affect disky pseudobulges. In fact, bar-driven secular evolution is the
primary process of disky pseudobulge creation and growth \citep{kk04, athanassoula05}.  Thus, the theoretical prediction is that
stronger bars push more gas inwards, resulting in more massive disky
pseudobulges. 

A visual approximation of bar formation and evolution is presented in Fig.~\ref{fig:bar_cartoon}.

\subsection{The Effect of Gas Content on Bar Formation\label{sub:pbar_gas}} 

We explain the trends we observe between the likelihood of disk galaxies being barred and their SSFR (present even at fixed $M_*$, $n$, or $\Sigma_{\rm1~kpc}^*$; see Fig.~\ref{fig:barfrac}) as being due to the effect of gas on bar formation. In the models, bars form later in disk galaxies with significant gas content, and after they form, they grow slower than disk galaxies with comparably less gas (AMR13 and Fig.~\ref{fig:barevol}). This predicts that the bar likelihood should be higher in gas-poor galaxies (i.e., the quiescent population), simply because some of the gas-rich galaxies (i.e., the star-forming sequence) have not yet formed their bars. Thus, taking SSFR as a tracer of gas content, then there is good agreement between simulation results and the trends we find \citep[see also][]{masters12}.

Within the star-forming sequence (defined here as $\log~\rm SSFR > -11~ \rm yr^{-1}$) disk galaxies do not all have the same $p_{\rm bar}$, but neither do they all have the same gas content. There are well known trends between SSFR, stellar mass, and gas content of disk galaxies \cite[e.g.,][]{catinella10, saintonge11}. The trend we observe here for $p_{\rm bar}$ to increase as SSFR declines (and $M_*$ increases) can be explained as being due to decreasing amounts of gas in the disks of these galaxies. Indeed, \cite{masters12} showed that if you correct for the typical HI content of a disk galaxy, those galaxies with more HI than is expected for their stellar mass are less likely to host bars. 

\subsection{The Effects of Classical Bulges and Disky Pseudobulges on Bar Formation}

We observe trends of bar likelihood with the S\'ersic index ($n$, Fig.~\ref{fig:barfrac}b) and central surface stellar mass density ($\Sigma_{\rm 1~kpc}^*$, Fig.~\ref{fig:barfrac}c), where these latter two parameters are considered to be measures of bulge prominence. In the star-forming sequence bar likelihood increases with both increasing $n$ and $\Sigma_{\rm 1~kpc}^*$, while the opposite trend is observed in the quiescent population. In order to interpret these trends, we need to remember that there are two main types of bulges -- the classical bulge and the disky pseudobulge, a distinction which will help explain this dichotomy.

\begin{figure*}[t!]  
\centering
\includegraphics[scale=.80]{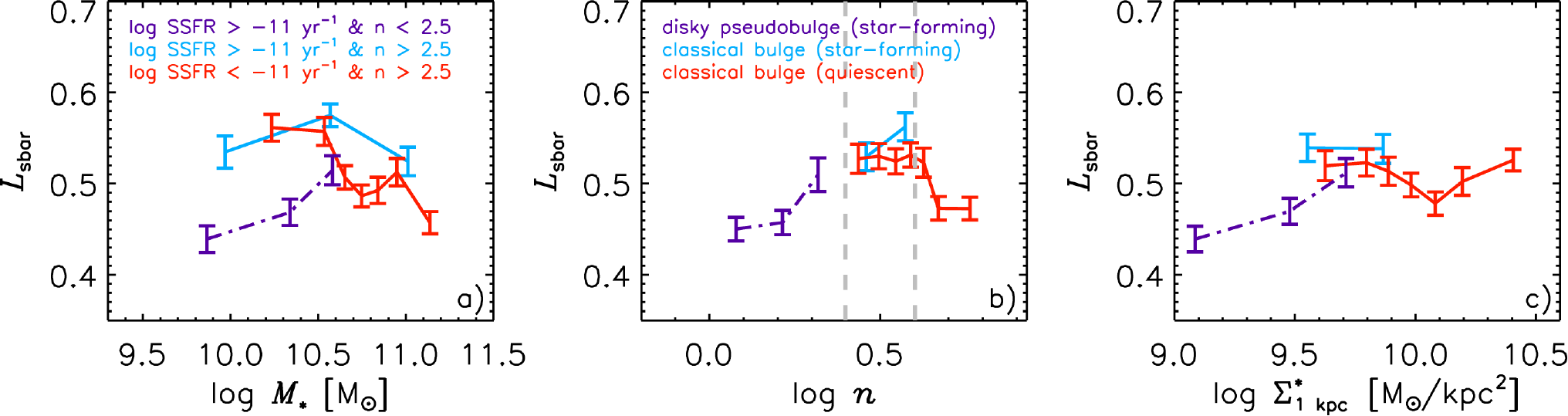}
\caption{Average $L_{\rm sbar}$ plotted against: {\it a)} $M_*$, {\it b)} $n$, and {\it c)} $\Sigma_{\rm 1~kpc}^*$. The details of this figure are identical to that of Fig.~\ref{fig:barlength_vs}, with the exception that each bin contains $\sim75$ galaxies and also, galaxies are further separated by bulge type, as identified by $n$. Purple points represent the star-forming disky pseudobulge galaxies, light blue points represent the star-forming classical bulge galaxies, and red points represent the quiescent classical bulge galaxies. The correlations of $L_{\rm sbar}$ with $n$ and $\Sigma_{\rm 1~kpc}^*$ for the disky pseudobulge galaxies match the predictions of bar-driven secular evolution. 
\label{fig:barlength_1d}}
\end{figure*}

The best way to distinguish these types of bulges involves the use of high resolution imaging of the bulges \citep[e.g.,][]{fisher08}, something that is not available for our large sample. However one can approximately separate the types with a threshold in the global galaxy S\'ersic index \citep{drory07}\footnote{\cite{gadotti09} advocates using the Kormendy relationship to separate classical bulges from disky pseudobulges. For this work, however, we choose to use the more simple global galaxy S\'ersic threshold.}. Disky pseudobulges generally lie in galaxies with global $n<2.5$ while classical bulges are found in galaxies with global $n>2.5$. Although this method is less accurate than those using high resolution imaging, this is a simple option that is adequate for our purposes. Hence, we adopt this S\'ersic threshold for the rest of the paper to distinguish the two types of bulges\footnote{Of course, there are $n<2.5$ galaxies that have no bulge \citep[e.g.,][]{simmons13}. However, for simplicity, we consider all galaxies with $n<2.5$ to contain a disky pseudobulge even if it might be a pure disk galaxy. This will not affect our discussion since pure disks and disky pseudobulges are closely related \citep[see][]{kk04}.}. This threshold is illustrated with a vertical dashed line at $\log~n=0.4$ ($n=2.5$) in Fig.~\ref{fig:barfrac}b. 
 
Our sample confirms the well known observation \citep[e.g.,][]{drory07} that quiescent (red) disk galaxies primarily have classical bulges while star-forming (blue) disk galaxies mainly have disky pseudobulges (see Fig. \ref{fig:barfrac}b).  This suggests that the decreasing $p_{\rm bar}$ with $n$ and $\Sigma_{\rm 1~kpc}^*$ observed in the quiescent disk galaxies is due to $p_{\rm bar}$ decreasing in galaxies with larger classical bulges, while the increasing $p_{\rm bar}$ with $n$ and $\Sigma_{\rm 1~kpc}^*$ observed in star-forming disks shows that $p_{\rm bar}$ is larger in galaxies with more massive disky pseudobulges.  

The classical bulge, like the halo, slows down bar formation due to it `diluting' the non-axisymmetric forcing of the bar \citep{athanassoula12}. This predicts that bar likelihood should decrease with increasing prominence of the classical bulge, as we indeed observe.

Disky pseudobulges result from the material that a bar pushes inwards to the central part of the
disk. Since these bulges formed after the bar, and are in fact, a product of the bar, they cannot
influence the bar formation phase. However there is a clear link
predicted between the existence of the bar and the amount of mass in
the disky pseudobulge, (or central 1 kpc; \citealt{athanassoula92,
  heller94, fisher10}). For galaxies of a given gas mass (or SSFR), a
higher bar likelihood should result in more massive disky
pseudobulges, as we observe (Figs.~\ref{fig:barfrac}b and
\ref{fig:barfrac}c). 

\subsection{Evidence for Secular Evolution} \label{sub:sec_evo}

Bar length trends (\S\ref{sub:barlength_results}) can help us
understand the secular evolution phase of the bar. We can safely 
assume that during the secular evolution phase of any
non-interacting galaxy its bar length may be considered a proxy of bar
age. However, this may not be true for any two galaxies, because
the galaxy with the youngest bar can have the longest bar, 
provided its halo can absorb larger parts of the angular momentum
emitted by the bar region (A03). Our comparisons, however, do not
concern two galaxies but ensembles of a relatively large number
of galaxies. For example in Fig.~\ref{fig:barlength_vs}b we
compare ensembles of galaxies     
with different $n$ values. But the number of galaxies in each
ensemble is sufficiently large for us to assume that galaxies
with a variety of halo properties are included in a roughly similar
manner in all ensembles. This subtle, but important point is intrinsic 
in our analysis and will be discussed further in \S\ref{sec:conclusions}. 

The strongest $L_{\rm sbar}$ trends we observe are found within star-forming disk galaxies (the average $L_{\rm sbar}$ increases monotonically with $M_*$, $n$, and $\Sigma_{\rm1~kpc}^*$; see Fig.~\ref{fig:barlength_vs}). To better understand the underlying physical processes responsible for these trends, we separate the data in Fig.~\ref{fig:barlength_vs} by bulge type; this is shown in Fig.~\ref{fig:barlength_1d}. Recall that galaxies with $n<2.5$ are considered to contain disky pseudobulges while galaxies with $n>2.5$ are considered to contain classical bulges. Note that quiescent galaxies with $n<2.5$ are very rare, hence they are not shown.

During the secular evolution phase, 
 bars become stronger, longer, and more efficient at funneling
  gas into the central regions of galaxies, leading 
  to more massive disky pseudobulges (AMR13). This
prediction matches our observations in Figs.~\ref{fig:barlength_1d}b
and \ref{fig:barlength_1d}c, where it is clear that $L_{\rm sbar}$ is
correlated with $n$ and $\Sigma_{\rm1~kpc}^*$ for the disky
pseudobulge galaxies (purple). These correlations give
evidence for the secular evolution phase of bars.  

The $L_{\rm sbar}$ trends with
  classical bulges are much more complex and also much less
  straightforward to interpret. 
  Simulations show that classical bulges should foster secular
  evolution by absorbing
  some of the angular momentum emitted by the bar region
  (A03). Hence the expectation is that galaxies with more massive classical
  bulges should have longer bars, but also longer disk scale
  lengths. 

Figs.~\ref{fig:barlength_1d}b and 
\ref{fig:barlength_1d}c show that the classical bulge galaxies (light blue and red) generally have longer scaled bar lengths than galaxies without a classical bulge, i.e., the disky pseudobulge galaxies. However, there is little evidence of increasing scaled bar length with increasing $n$ and $\Sigma_{\rm~1 kpc}^*$. In fact, there actually appears to be a decrease in scaled bar length for $\log~n$ larger than 0.6 (i.e., $n$ larger than about 4, equivalent to a more
concentrated light profile than the standard $r^{1/4}$ de Vaucouleurs profile). Similarly the scaled bar length stops increasing in the rightmost panel for 
$\Sigma^*_{\rm1 kpc}$ larger than roughly $10^{9.8}~
\rm{M_{\odot}/kpc}^2$ (although there seems to be an final increase in scaled bar length at the highest $\Sigma_{\rm~1 kpc}^*$).

%Figs.~\ref{fig:barlength_1d}b and 
%\ref{fig:barlength_1d}c argue that up to $\log~n$ of the order of 0.6 ($n\approx4$; marked by a 
%vertical dashed line), and for $\Sigma^*_{\rm1~kpc}$ roughly up to 10$^{9.8}$ M$_{\odot}/\rm kpc^2$ the bar length increases faster than the disk scale length, so that
%their ratio, i.e., $L_{\rm sbar}$,
% is an increasing function of both $\log~n$ and 
%$\Sigma^*_{\rm1~kpc}$. Also, for these values, 
%classical bulge galaxies, both star-forming (light blue)
%and quiescent (red), generally contain bars with larger $L_{\rm sbar}$
%than those of the disky pseudobulge galaxies (purple).
%
%These two panels show a further interesting feature. Namely for
%$\log~n$ larger than 0.6 (i.e., $n$ larger than about 4, equivalent to a more
%concentrated light profile than the standard $r^{1/4}$ de Vaucouleurs profile), the scaled bar 
%length {\it decreases} with increasing global S\'ersic index.
%Similarly the scaled bar length stops increasing in the rightmost panel for 
%$\Sigma^*_{\rm1 kpc}$ larger than roughly $10^{9.8}~
%\rm{M_{\odot}/kpc}^2$.

This decrease in scaled bar length with large $n$ and $\Sigma_{\rm 1~kpc}^*$ does not disagree with simulation results, and
can be attributed to the presence of a very high central mass
concentration {(CMC)}\footnote{We note that this is just one of the possible reasons for this observed decrease. This could also correspond to the regime where the bulge is so massive that it has significantly delayed the onset of bar formation, resulting in a lack of bar evolution.}. Indeed, our last averaged point is roughly at a $\log~n$ value of
0.8, which corresponds to a S\'ersic index of roughly 6.5. This could 
well be due to a luminosity spike in the center of the galaxy which
would hamper the bar growth and evolution if it pre-existed the bar,
or if grown later, that would 
bring a decrease of the bar length and strength
\citep[e.g.,][]{shen04, athanassoulaLD05}. This strong 
CMC will thus bring a decrease of bar length at the highest values of  $n$, 
as seen in Fig.~\ref{fig:barlength_1d}b (and Fig.~\ref{fig:barlength_vs}b).  

Nevertheless, at least part of this decrease could be spurious and
  due to the fact that the bar component is not specifically included
  in our 2-component decompositions, which is more
  worrisome for galaxies with stronger and longer bars. To test it we
  scaled the {\it absolute} bar length with the $r$-band isophotal radius at 
  25 mag arcsec$^{-2}$ from the SDSS pipeline and re-created
  Figs.~\ref{fig:barlength_vs} and \ref{fig:barlength_1d}. The results can be found in Appendix~\ref{appendix:isoa}. 
  We find then that the decrease seen with the 
  disk scale length at high $n$ and $\Sigma_{\rm 1~kpc}^*$ 
  is considerably lessened. We do not fully understand the bar length trends with the classical bulge galaxies at the highest $n$ and $\Sigma_{\rm 1~kpc}^*$, more work needs to be done.
  Let us note, however, that
  the correlations of the scaled bar length with the disky pseudobulge galaxies
  are still present even when scaling with
  the isophotal radius, thus enhancing our confidence in the
  corresponding decompositions and trends. 

\section{Discussion}

\subsection{Are We Observing Secular Evolution?}

In \S\ref{sec:results} and \S\ref{sec:compare_theory}, we showed evidence which suggests that disky pseudobulges are more massive in populations of galaxies which are more likely to host bars and which host longer bars (specifically that average values of $p_{\rm bar}$ and $L_{\rm sbar}$ increased with $n$ and $\Sigma_{\rm1~kpc}^*$ for disky pseudobulge galaxies). We interpreted this as observational evidence of bar-driven secular evolution growing disky pseudobulges \citep{kk04, athanassoula05}. Our interpretation hinges on the assumption that bar length traces the evolution of bars. This assumption is based on both simulations of bar growth as well as observational data. \cite{elmegreen07} showed that bar length mirrors bar strength \citep[see also][]{block04}. The
simulations of bar growth shown in Fig.~\ref{fig:barevol} -- and a
large number of others, as reviewed by \cite{athanassoula12} -- demonstrate that
isolated bars typically grow stronger with time. 

Furthermore, the simulations of AMR13 argue that bars in isolated
galaxies are long-lived structures -- in the $\sim10$ Gyrs that their
simulations covered, not one of their bars dissolved (see also
\citealt{debattista06} and \citealt{berentzen07} for a similar
conclusion). Recent zoom-in cosmological simulations by
\cite{kraljic12} also support the idea that bars are long-lived
structures. Their simulations show that most of the bars that formed
at $z\le1$, when mergers have become less frequent, persist down to $z=0$. Observational studies have now observed bars with modest frequencies out to $z\sim1$ (\citealt{abraham99, jogee04, elmegreen04, sheth08, cameron10, melvin13}), and one upcoming study detects bars as far out as $z\sim1.5$ (J. Herrington et al. 2013, submitted.), with the implication that many of the bars we observe in the local Universe could have formed at $z\sim1$ or earlier. This gives a substantial time window for secular evolution to grow longer bars and stronger disky pseudobulges.

Previous works have shown a trend between bar length and Hubble type -- that bars are longer in earlier type disks -- and used this to argue that secular evolution had been observed \citep[e.g.,][]{athanassoula80, elmegreen85, martin95, regan97, erwin05, laurikainen07, elmegreen07, menendez07, gadotti11}. Our result is novel in that it looked for trends of bar length with the central mass density in the very centers of galaxies, a quantity that is directly linked to secular evolution in models. Our sample is also nearly an order of magnitude larger than any previous study. Thus, we argue that our result is the best evidence yet for bar-driven secular evolution in disk galaxies. 

Recent results from several high resolution simulations present mechanisms for the formation of disky pseudobulges that do not rely on secular evolution \citep{inoue12, okamoto13}, but rather involve dynamical instability in clumpy galaxies or high-redshift starbursts.  While the bulges of these simulations do have characteristics of local disky pseudobulges\footnote{Not all bulges made from clump coalescence have characteristics of disky pseudobulges. For example, \cite{elmegreen08} show that their bulges made through clump coalescence have properties of classical bulges.}, our results here suggest that secular evolution does have a major effect, both in creating disky
pseudobulges and in building up the stellar mass in the bulge
region of barred galaxies. 

Nevertheless, there are substantial numbers of disk galaxies that are
non-barred and are hosting disky pseudobulges \citep{kk04}. Up to a
third or more of the local disk galaxy population is unbarred in even
the most conservative reckoning. This observations argues that disky pseudobulges
have more than one formation mechanism. Perhaps disky pseudobulges in
non-barred galaxies were created through high redshift channels, while
the disky pseudobulges in barred galaxies may have been created, and
are still in the process of growing through bar-driven
secular evolution, at much later times.

\subsection{Can Bars Quench Star Formation?} 

The highest values of $p_{\rm bar}$ are found among quiescent galaxies with $n\sim2.5$ (see Fig.~\ref{fig:barfrac}b). Here we consider the question of whether these bars were formed in situ or if they could be implicated in the processes which turned these disk galaxies quiescent. We ask `were these bars formed in $n\sim2.5$ quiescent galaxies, or did they form in star-forming galaxies (with $n\ls2.5$) that evolved into the $n\sim2.5$ quiescent disk galaxies?' We refer to this latter process as `bar quenching' and explore this idea further.  

Bars have been associated with enhanced central star formation in galaxies for decades \citep{hawarden86, dressel88, friedli93, giuricin94, huang96, martinet97, martin97, ho97, ellison11, oh12,  wang12}. This is a natural consequence of the evolution of gas in a disk galaxy under the influence of a bar. The bar-induced gravitational torques funnel gas into the centers of galaxies \citep{matsuda77, simkin80, athanassoula92, wada&habe92, wada&habe95, friedli93, heller94, knapen95, sakamoto99, sheth05}, where it should quickly form stars, thus enhancing the central star formation. If this secular evolution were efficient, it could accelerate the depletion of the gas supply within a considerable fraction of the disk, namely the region within corotation. If this process were not balanced by an increased inflow of cosmological gas, this would ultimately, produce a quiescent barred galaxy \citep{masters11,masters12}. 

Large surveys such as SDSS \citep{abazajian09}, COSMOS \citep{scoville07}, and AEGIS \citep{davis07} have painted a clear picture of the structural properties of quiescent galaxies -- they are massive, centrally concentrated, and have high central velocity dispersions \citep[e.g.,][]{franx08, bell12, wake12, cheung12, barro13, fang13, williams13}. \cite{cheung12} recently found that the most distinguishing structural parameter of quiescent galaxies (compared to star-forming galaxies) is their central surface stellar mass density (within a radius of 1 kpc). Almost all quiescent galaxies in the sample of \cite{cheung12} have high values of $\Sigma_{\rm1~kpc}^*$, while star-forming galaxies mainly have low values of $\Sigma_{\rm1~kpc}^*$. This is clear evidence that the process(es) that quench star formation in these galaxies are related to the buildup of the central stellar mass density \citep[see also][]{fang13}. We consider here if secular evolution is {\it able} to build high enough central densities to act as a quenching mechanism.

Indeed, Fig.~\ref{fig:barlength_1d}c shows that the $\Sigma_{\rm 1~kpc}^*$ values of the disky pseudobulge galaxies overlap partly with the $\Sigma_{\rm1~kpc}^*$ values of the classical bulge galaxies. The most massive of the disky pseudobulges we argue are grown by bar-driven secular evolution are comparable in central density to the smallest of the classical bulges. This suggests that secular evolution {\it can} build the high central densities that are observed in quiescent galaxies. This appears to be circumstantial evidence for an interesting, and potentially important galaxy evolution process -- bar quenching. We caution, however, that our identification of disky pseudobulges in quiescent barred disk galaxies is based on global S\'ersic fits. If it is indeed the case that there exist quiescent disk galaxies which host only disky pseudobulges, and show no evidence for classical bulges, this will be strong evidence for the process of 'bar quenching' having acted in these galaxies. However, more accurate identifications of disky pseudobulges are needed to verify this claim.

\section{Conclusion} \label{sec:conclusions}

In this paper, we use hundreds of thousands of visual classifications measurements of galactic bars provided by ``citizen scientists'' through the Galaxy Zoo project \citep{lintott08, lintott11, willett13}. We first select a sample of disk galaxies in which reliable bar classifications can be made -- we call this the Galaxy Zoo 2 Disk (GZ2D) sample, which comprises 13,295 oblique (i.e., face-on or mildly inclined) disk galaxies in a volume limit to $z=0.06$. This sample is similar to the GZ2 samples used previously to study trends of the bar fraction by \cite{masters11} and \cite{masters12}. Strongly barred galaxies identified in GZ2 were part of a small Galaxy Zoo project which used a Google Sky interface to collect measurements of bar lengths \citep{hoyle11}. In this paper we also make use of this Bar Length (BL) sample, which comprises 1,154 galaxies. We use these data to analyze the dependence of bar likelihood ($p_{\rm bar}$, a weighted and debiased fraction of GZ users identifying a bar, and which acts like a probability of a galaxy containing a visually identifiable bar; \citealt{willett13}) and scaled bar length ($L_{\rm sbar} = L_{\rm bar}/2.2 r_d$; a measure of bar strength, linked to how evolved a bar is) on other galactic properties. Specifically we test how the likelihood and length of bars depend on specific star formation rate (SSFR; estimated through nebular emission lines from the SDSS fiber, and the broadband {\it ugriz} SDSS photometry, as measured by MPA-JHU) and inner galactic structure (i.e., bulge prominence) parameterized by global S\'ersic index, $n$, as measured by GIM2D, and central surface stellar mass density, $\Sigma_{\rm1~kpc}^*$, as estimated from a 1 kpc radius circular aperture projected onto SDSS images.

Our main observational results (\S\ref{sec:results}) are:

\begin{enumerate}

\item There exists an anti-correlation between $p_{\rm bar}$ and SSFR; this relationship is present even at fixed $M_*$, $n$, or $\Sigma_{\rm 1~kpc}^*$.

\item The structural trends of $p_{\rm bar}$ are bimodal with SSFR. In the star-forming sequence, $p_{\rm bar}$ correlates with $n$ and $\Sigma_{\rm 1~kpc}^*$, while in the quiescent population, $p_{\rm bar}$ anti-correlates with $n$ and $\Sigma_{\rm1~kpc}^*$. 

\item The structural trends of $L_{\rm sbar}$ are also bimodal with SSFR. Within the star-forming sequence, $L_{\rm sbar}$ correlates with $n$ and $\Sigma_{\rm1~kpc}^*$, in a similar way to $p_{\rm bar}$. However within the quiescent population, $L_{\rm sbar}$ shows a rather different behavior, with a peak at values of $n\sim4$ and $\Sigma_{\rm1~kpc}^* \sim 10^{9.8}$ M$_\odot$/kpc$^{2}$. 

\end{enumerate}

We compare these results to simulations of bar formation and evolution in \S\ref{sec:compare_theory}. We find that the underlying physical processes become clearer upon separating these galaxies by those that contain disky pseudobulges ($n<2.5$) and those that contain classical bulges ($n>2.5$). This comparison reveals the following:

\begin{enumerate}

\item Assuming that SSFR is a good tracer of gas content, the anti-correlation of $p_{\rm bar}$ with SSFR is consistent with the expected effects of gas on {\it bar formation}. Simulations show that gas delays the formation of bars, thus many gas-rich galaxies simply have not yet formed bars, while most gas-poor disk galaxies have.

\item The observed trends of $p_{\rm bar}$ and
  $L_{\rm sbar}$ with $n$ and $\Sigma_{\rm 1~kpc}^*$ for classical
  bulge galaxies are consistent with the effects of classical bulges
  and CMCs on bar formation and evolution. The gravitational forcing
  of classical bulges `dilute' the non-axisymmetric forcing of the
  bar, which delays the formation of a bar. This diluting effect is
  more powerful in more massive classical bulges, resulting in a
  longer delay of {\it bar formation}. After the bar has formed,
  however, classical bulges are expected to promote {\it
    secular evolution} by absorbing the angular momentum
  emitted from the bar region; this process also scales with the mass of the
   bulge and leads to both longer bars and longer disk scale
    lengths. Our results suggest that for S\'ersic index up to roughly $n=4$ the bar
    length may increase faster than the disk scale length. For yet higher
    values of $n$, a strong ensuing CMC could
    lead to a decrease of the bar strength by generating instabilities of 
  the main family of bar-supporting orbits. 
    
\item The correlations of $p_{\rm bar}$ and $L_{\rm sbar}$ with $n$ and $\Sigma_{\rm 1~kpc}^*$ for the disky pseudobulge galaxies are in agreement with the predictions of {\it bar-driven secular evolution}. Bars drive gas toward the centers of galaxies, where the gas should eventually form stars and give rise to disky pseudobulges. As bars grow stronger and longer, the ability to funnel gas grows stronger as well, resulting in more massive disky pseudobulges. 

\end{enumerate}

The comparison of the observational results we present here with simulations of bar formation and growth shows general agreement, indicating that many of the underlying physical processes of bar formation and evolution are understood. An implication of this is that we are confident in our basic understanding of the relationship between bars and their host galaxies. Bars are clearly not stagnant structures, rather they are dynamic, evolving, and furthermore directly influence the evolution of their host galaxies. 

While this work only concerns the universe at $z\sim0$, the ramifications of this idea reach far beyond the local universe. There is increasing evidence that bars have been present since $z\sim1$ (\citealt{abraham99, jogee04, elmegreen04, sheth08, cameron10, melvin13}; J. Herrington et al. 2013, submitted.), indicating that the evolution of disk galaxies has been affected by bars for the last $\sim 8$ billion years. Moreover, if the observed evolution of bar fraction with redshift is extrapolated into the future (there is now an agreement that bar fraction increases towards lower $z$), then bars will soon be present in nearly all disk galaxies, and hence become an even more dominant driver of disk galaxy evolution. 
  
We can not yet claim to understand all aspects of the symbiotic
relationship between bars and their host galaxies. We do not fully
understand the complicated behavior we observe between bar length and
inner galactic structure in disk galaxies hosting classical
bulges. Our tentative explanation is that these trends are due to the
presence of CMCs, however this should
be tested with much higher resolution imaging to probe the very
centers of galaxies. Furthermore, in this work we have not explored
many of the parameters that are predicted to affect bar formation and
evolution (e.g., the dark matter halo and the velocity dispersion of
the stars in the disk, \citealt{Athanassoula.Sellwood.86};
A03). Even so, we found a good
agreement between theory and observations and all observational
trends could be well explained by simulations. This may be due to the
large size of our sample, which allows for a variety of halo 
properties and of disk velocity dispersions in a roughly similar
manner in all ensembles we compared. Finally, the role bars may play in processes which quench star formation is an interesting, and potentially important issue for galaxy evolution that warrants further study. 

The most notable success in our comparison between observation and theory is the evidence we present for secular evolution. Unlike galaxy mergers, secular evolution is a slow and gentle process that is not immediately obvious in images. There has been previous observational evidence of secular evolution in galaxies \citep[e.g.,][]{athanassoula80, courteau96, macarthur03, elmegreen07, laurikainen09, coelho11, sanchezjanssen13}, however the combination of our large dataset and the observed correlations of bar likelihood {\it and} bar length with inner galactic structure for star-forming disk galaxies makes our results one of the most compelling pieces of evidence of not only the existence of secular evolution, but also of the role of ongoing secular processes on the evolution of disk galaxies. 

\acknowledgments

This publication has been made possible by the participation of more than 200000 volunteers in the Galaxy Zoo project. Their contributions are individually acknowledged at http://www.galaxyzoo.org/Volunteers.aspx. 

EC acknowledges financial support from the National Science Foundation Grant AST 08-08133. EA and AB acknowledge financial support to the DAGAL network from the People Programme (Marie Curie Actions) of the European Union's Seventh Framework Programme FP7/2007-2013/ under REA grant agreement number PITN-GA-2011-289313. They also acknowledge financial support from the CNES (Centre National d'Etudes Spatiales - France). KLM acknowledges funding from The Leverhulme Trust as a 2010 Early Career Fellow. RCN acknowledges STFC Rolling Grant ST/I001204/1 "Survey Cosmology and Astrophysics". KS gratefully acknowledges support from Swiss National Science Foundation Grant PP00P2\_138979/1. RAS is supported by the NSF grant AST-1055081. KWW is supported by the US National Science Foundation under grant DRL-0941610. EA, EB and AB thank the Aspen Center for Physics for their hospitality during the workshop "The Milky Way as a Laboratory for Galaxy Formation" in 2013 and the NSF for partial financial support. Part of the simulation work described here was performed using HPC resources from GENCI- TGCC/CINES (Grant 2013 - x2013047098).

Funding for the SDSS and SDSS-II has been provided by the Alfred P. Sloan Foundation, the Participating Institutions, the National Science Foundation, the U.S. Department of Energy, the National Aeronautics and Space Administration, the Japanese Monbukagakusho, the Max Planck Society, and the Higher Education Funding Council for England. The SDSS website is http://www.sdss.org/. The SDSS is managed by the Astrophysical Research Consortium for the Participating Institutions. The Participating Institutions are the American Museum of Natural History, Astrophysical Institute Potsdam, University of Basel, University of Cambridge, Case Western Reserve University, University of Chicago, Drexel University, Fermilab, the Institute for Advanced Study, the Japan Participation Group, Johns Hopkins University, the Joint Institute for Nuclear Astrophysics, the Kavli Institute for Particle Astrophysics and Cosmology, the Korean Scientist Group, the Chinese Academy of Sciences (LAMOST), Los Alamos National Laboratory, the Max-Planck-Institute for Astronomy (MPIA), the Max-Planck-Institute for Astrophysics (MPA), New Mexico State University, Ohio State University, University of Pittsburgh, University of Portsmouth, Princeton University, the United States Naval Observatory, and the University of Washington.

EC would like to thank Charlie Conroy, Jonathan R. Trump, Guillermo Barro, Jerome Fang, and Yicheng Guo for useful discussions, comments, and suggestions. We also thank the anonymous referee for a very helpful report that resulted in substantial improvements to the paper.

\appendix

\section{Completeness} \label{appendix:completeness}

\begin{figure*}[t!] 
\centering
\includegraphics[scale=.6]{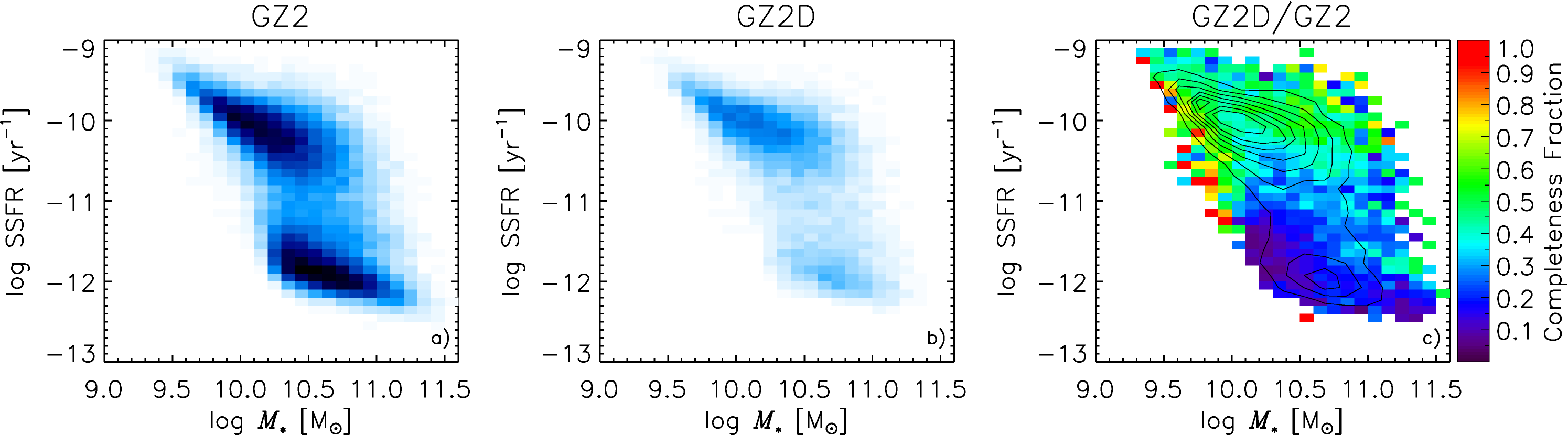}
\caption{The number density distribution of the: {\it a)} volume-limited parent Galaxy Zoo 2 (GZ2) sample and {\it b)} Galaxy Zoo 2 Disk (GZ2D) sample; both are scaled to the same. {\it c)}: The completeness of the GZ2D sample relative to the GZ2 sample. For each bin, we calculate the fraction of GZ2D galaxies in the GZ2 sample and color it according to the color bar to the right. The black contours outline the number density of the GZ2D sample and only bins with at least 2 GZ2 galaxies are shown. The completeness of GZ2D is bimodal such that it recovers $\sim50\%$ of high SSFR ($>10^{-11}~ \rm yr^{-1}$) galaxies and $\sim20\%$ of low SSFR ($<10^{-11}~ \rm yr^{-1}$) galaxies.
\label{fig:completeness}}
\end{figure*}

\begin{figure}[t!] 
\centering
\includegraphics[scale=.6]{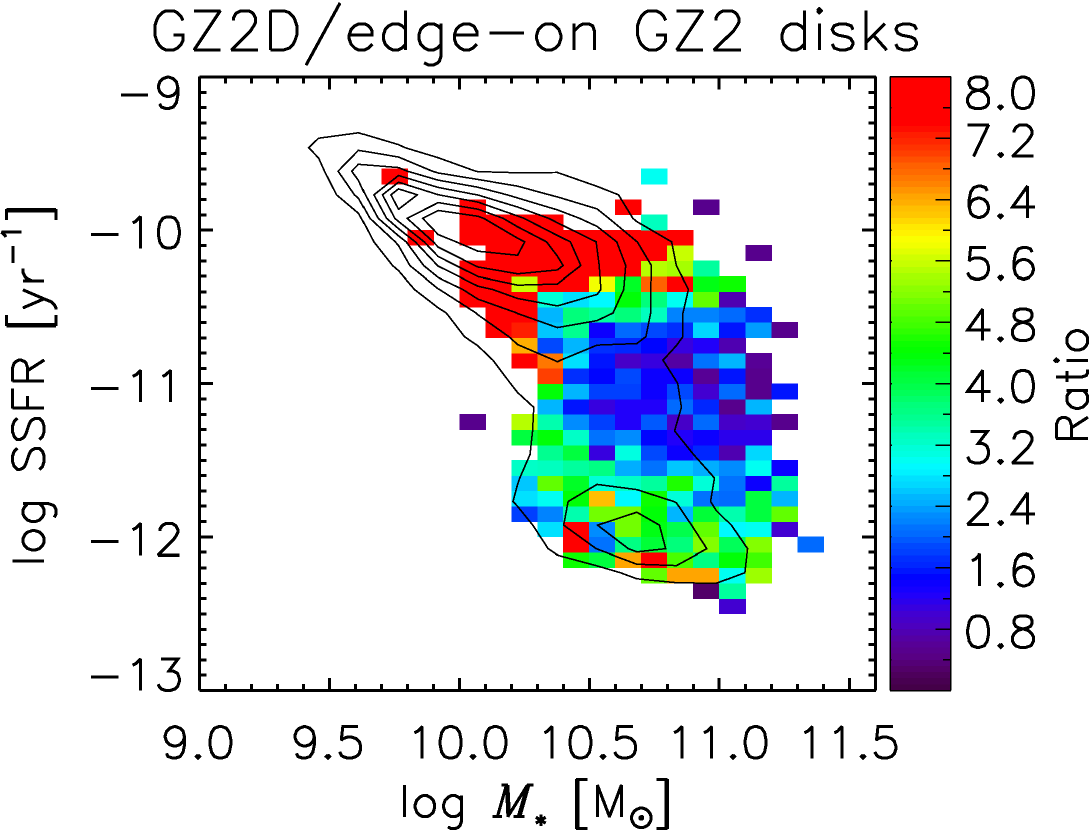}
\caption{The ratio of the number of galaxies in the GZ2D sample to the edge-on GZ2 disk sample. The black contours outline the number density distribution of the GZ2D sample. Only bins with $n \ge 2$ edge-on GZ2 galaxies are shown. GZ2D does not seem to be strongly biased against low mass quiescent disks. 
\label{fig:completeness_edge}}
\end{figure}

\subsection{Galaxy Zoo 2 Disk Sample} \label{sub:comp_gz2}

The Galaxy Zoo 2 Disk (GZ2D) sample was selected on the basis of hundreds of thousands of visual morphological classifications collected via the Galaxy Zoo website. In order for a galaxy to be in this sample, the majority of volunteers classifying it must have identified `features' in it, and identified it as not being an edge-on disk \citep[see][]{masters11, willett13}. In addition we apply an extra cut at $b/a>0.5$ to ensure disks are face-on enough to identify bars. This results in a sample of disk galaxies with a broad mix in Hubble types. Objects might contain an obvious disk (e.g., SBc, Sc) or a subtle disk (e.g., S0). While we do not expect problems in identifying the former in any orientation, S0 galaxies are notoriously difficult (even for the most expert classifiers) to separate from ellipticals, if viewed face-on. We consider in this section if any face-on disk galaxies are missing from our sample. Presumably, if a galaxy had a bar, it would be readily identify as `featured' and included in this sample, thus we assume any missing disk galaxies will be non-barred and therefore introduce potential biases into our results. 

We use as a comparison sample, the volume-limited parent Galaxy Zoo 2 sample (see \S\ref{data:gz} for details of our initial Galaxy Zoo 2 sample) that the GZ2D sample was originally drawn from, as well as a sample of edge-on disk galaxies in which we expect all types of disks will be equally easy to identify. As a reminder, the parent Galaxy Zoo 2 sample has the following criteria:

\begin{enumerate*}
\item $0.01 < z < 0.06$, where $z$ is the SDSS spectroscopic redshift.
\item $M_{r} \le -20.15$, where $M_{r}$ is the rest-frame absolute Petrosian $r$-band magnitude.
\end {enumerate*}

Hereafter, this sample will be referred to as the Galaxy Zoo 2 sample, or simply, the GZ2 sample. We match the GZ2 sample to the MPA-JHU catalog for stellar masses and star formation rates, resulting in a total of 43,221 galaxies. 

To identify edge-on disks, we use thresholds in the Galaxy Zoo vote fractions for `features of disk' ($p_{\rm features} > 0.5$) and for `edge-on disk' ($p_{\rm edge-on} > 0.80$), this is slightly more conservative that the recommended thresholds for selecting a `clean edge-on' sample as given in \cite{willett13}, but we do not expect the selection to introduce any bias with Hubble type for disk galaxies. 

Fig.~\ref{fig:completeness} compares the number density distribution of the GZ2D sample (panel b) to that of the whole volume limited GZ2 sample (panel a). Both panels are scaled so that the blue scale indicates the same range of density and only bins with at least 2 galaxies are shown. Using the GZ2 sample as the fiducial completeness standard, panel c displays the completeness of the GZ2D sample (i.e., the fraction of the GZ2 sample which is in GZ2D) as indicated by the legend. To aid the eye, contours of the GZ2D number density distribution are over-plotted. We point out that completeness levels of greater than $50\%$ are not expected since the selection on axial ratio ($b/a>0.5$) removes approximately half of all disk galaxies. In this plot we observe the expected bimodality, such that the completeness of high SSFR ($> 10^{-11}~ \rm yr^{-1}$; `star-forming') galaxies is much higher ($\approx 50\%$ complete) than it is for low SSFR ($ < 10^{-11}~ \rm yr^{-1}$; `quiescent') galaxies ($\approx 20\%$ complete). This reveals the well know correlation between SSFR and morphology - that most star-forming galaxies have disks, and many quiescent galaxies are elliptical, so do not have obvious `features' to be selected as part of the GZ2D sample.

This test, however, cannot reveal if the GZ2D sample represents a fair selection of all disks. To test that, we isolate a sample of edge-on disk galaxies in which we expect all disks (even S0s) will be identified. If the GZ2D sample is fairly representative of all disk galaxies, then the ratio of GZ2D (face-on disk) galaxies to the sample of edge-on disks should be uniform throughout the SSFR-mass diagram (this assumes all disk galaxies are randomly orientated, which we expect they should be, but also see \citealt{simard11}, and that the inclination introduces no systematic biases into estimates of SSFR or stellar mass, which is less clear). 

Fig.~\ref{fig:completeness_edge} compares the number density of the edge-on GZ2 disks to our GZ2D sample of mildy inclined or face-on disk galaxies.  We show the ratio of the number of galaxies in the GZ2D sample to the edge-on GZ2 disks sample. Only bins with at least 2 galaxies from the edge-on GZ2 disks sample are shown and the black contours represent the number density distribution of the GZ2D sample. For high SSFR galaxies, there are $\sim$7 galaxies in the GZ2D sample for every edge-on GZ2 disk galaxy. This is likely due to a combination of the expected number ratios for edge-on and not edge-on disk galaxies (e.g., for random orientations, we expect one galaxy with $i>85^\circ$ for every five with $i<65^\circ$), and the possible effects of increased internal extinction in the egde-on sample causing SSFR to be underestimated. However we do not expect to be missing systematically any star-forming disk galaxies. 

Because of the extinction of edge-on galaxies, the sample of low SSFR ($\log ~\rm SSFR < -11.6 ~ \rm yr^{-1}$) edge-on GZ2 disks may contain a combination of truly low SSFR disks and reddened intermediate SSFR disks. However, we assume that the reddened intermediate SSFR contribution to the low SSFR regime of the edge-on GZ2 disks sample is uniform across stellar mass and only changes the absolute scaling of the number ratio between GZ2D and edge-on GZ2 disks. We therefore examine the uniformity of the low SSFR regime in Fig.~\ref{fig:completeness_edge} to gauge whether the GZ2D sample is missing any quiescent disks. 

The number ratio between edge-on quiescent disks and face-on quiescent disks is largely uniform (at $\sim 5$ oblique disks per edge-on disk). There is, however, hints of a small dearth in the GZ2D sample at low masses. Averaging the number ratios at low masses ($\log ~M_* < 10.6$) reveals that we find $\approx10\%$ less GZ2D galaxies compared to the average number ratios of the high mass quiescent disks. The total number of low mass quiescent galaxies in our sample is $\approx900$, so this suggests we may be missing $\approx90$ low mass quiescent disk galaxies. Presumably, if a galaxy had a bar, it would be readily spotted and included in this sample, thus we assume the missing disk galaxies are non-barred. We assume that the missing disks have values of $n$ and $\Sigma_{\rm 1~kpc}^*$ typical for GZ2D galaxies of the same mass and SSFR. We find that the $n$ and $\Sigma_{\rm 1 ~kpc}^*$ values of these low mass quiescent disks are roughly uniformly distributed, meaning that the $p_{\rm bar}$ trends with $n$ and $\Sigma_{\rm 1~kpc}^*$ for the quiescent population are unaffected by this incompleteness. 
We can estimate how many unbarred quiescent disk galaxies we are missing for every bin by simply dividing the total number of missing galaxies ($\approx90$) by the total number of bins that the quiescent population spans in $n$ and $\Sigma_{\rm 1~kpc}^*$, which turns out to be $\sim20$ bins.
 Thus we are missing $\approx5$ unbarred quiescent disks in every $N=100$ bin of $n$ and $\Sigma_{\rm 1~kpc}^*$ (see Figs.~\ref{fig:barfrac}b and \ref{fig:barfrac}c). Even if all five galaxies have $p_{\rm bar} = 0$ this would reduce the average $p_{\rm bar}$ in each bin by at most 5\% (by simply adding 5 more galaxies in the denominator).

The number of missing low mass disks in the affected part of Fig.~\ref{fig:barfrac}a works out to be $\approx10$ per low mass quiescent bin (there are $\sim10$ bins in the low mass quiescent regime). Fig.~\ref{fig:barfrac}a shows that the $p_{\rm bar}$ values for the low mass quiescent bins are $\sim0.60$. Adding 10 non-barred ($p_{\rm bar}=0$) disk galaxies to these bins, i.e., adding 10 galaxies to the denominator, reduces these $p_{\rm bar}$ values to $\sim0.50$. Our qualitative results and interpretation are unaffected. Therefore, the missing non-barred low mass quiescent disks do not significantly influence our results. 

\subsection{Bar Length Sample} \label{sub:comp_bl}

\begin{figure}[t!] 
\centering
\includegraphics[scale=.6]{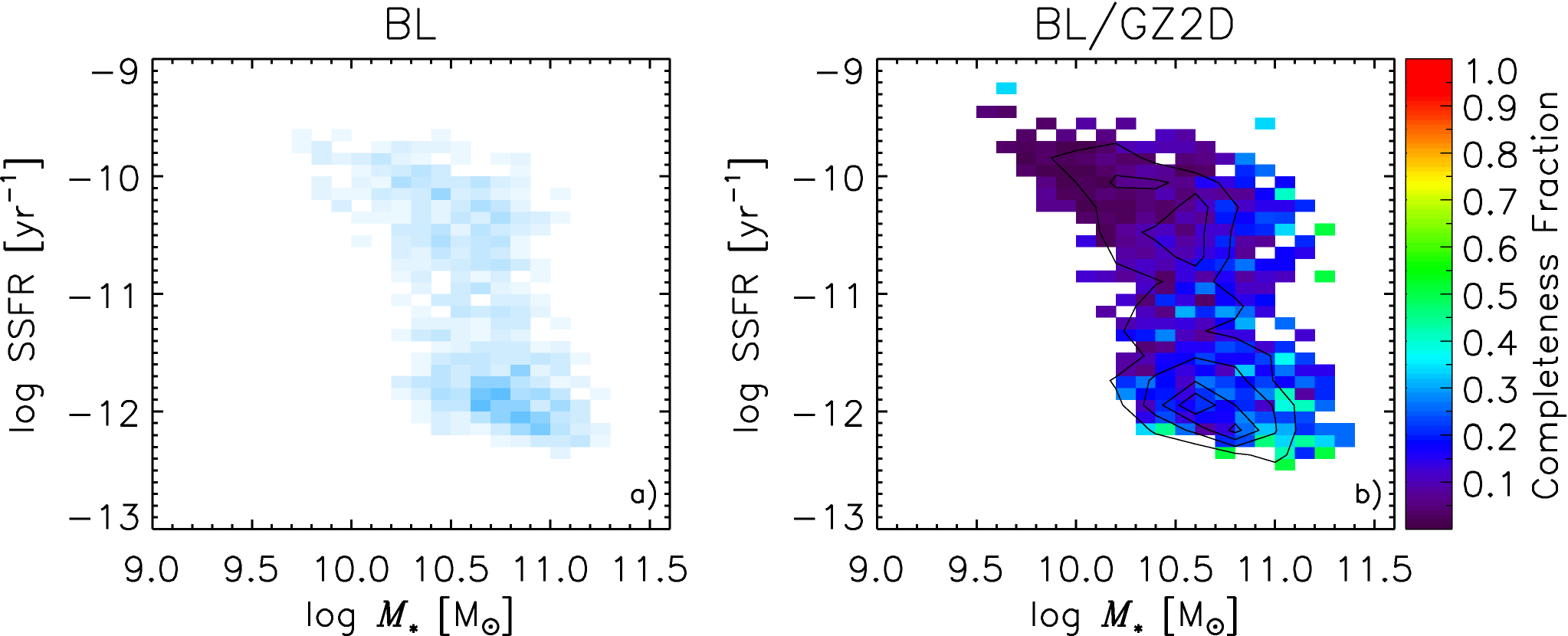}
\caption{{\it a)}: The number density distribution of the Bar Length (BL) sample scaled to a quarter of panel a of Fig 8. {\it b)}: The completeness of the BL sample relative to the GZ2D sample; the black contours in this panel represents the number density distribution of the BL sample. Only bins with 2 or more GZ2D galaxies are shown.
\label{fig:completenessBL}}
\end{figure}

Unlike the GZ2D sample, there is not a concern that the BL sample is missing non-barred disks since, as the sample name implies, the BL (Bar Length) sample only contains barred disks. Nevertheless, we want to ensure that it is not suffering any selection bias.  

Fig.~\ref{fig:completenessBL} shows the completeness of the BL sample relative to the GZ2D sample. The completeness of BL is approximately bimodal with SSFR. In the high SSFR regime ($\log ~\rm SSFR > -11 ~ \rm yr^{-1}$), the BL sample is $\sim10\%$ complete, while in the low SSFR regime ($\log ~\rm SSFR < -11 ~ \rm yr^{-1}$), it is $\sim20\%$. 

This completeness bimodality is reasonable because the BL sample is primarily composed of strong bars, which as illustrated in Fig.~\ref{fig:barfrac}, strong bars mainly lie in the quiescent population. However, since our analysis splits the BL sample into star-forming and quiescent (i.e., Fig.~\ref{fig:barlength_vs} and Fig.~\ref{fig:barlength_1d}), this difference in completeness should be inconsequential to our results and interpretations. 

\section{Bar Length Scaled by Isophotal Radii} \label{appendix:isoa}

\begin{figure*}[t!]  
\centering
\includegraphics[scale=.8]{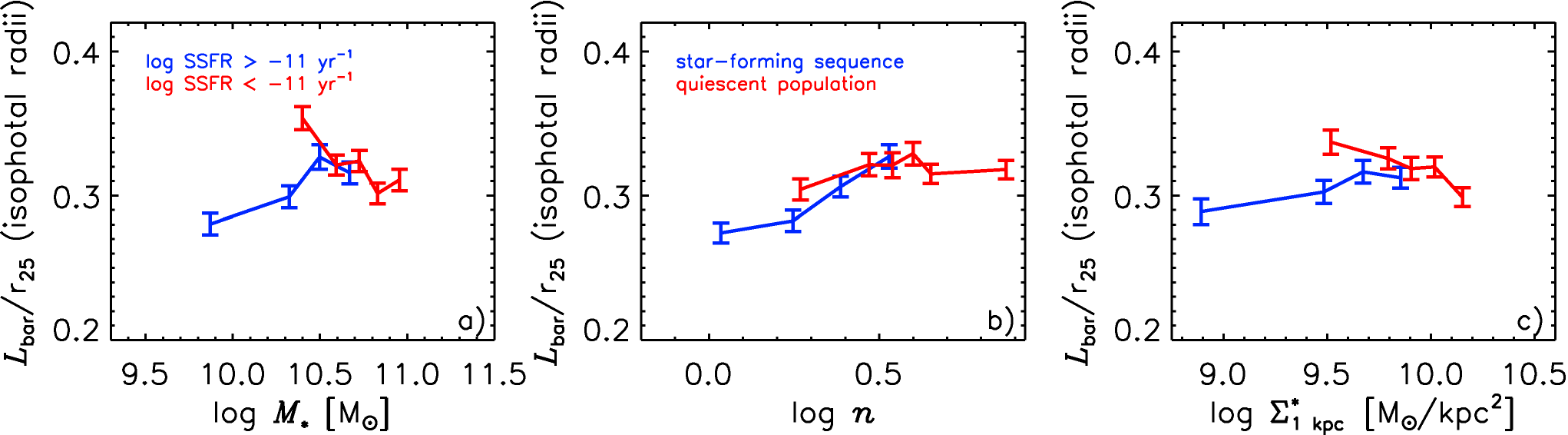}
\caption{Average bar length scaled by the isophotal radii, $L_{\rm bar}/\rm{r_{25}}$, plotted against: {\it a)} $M_*$, {\it b)} $n$, and {\it c)} $\Sigma_{\rm 1~kpc}^*$. Galaxies were split by their star formation state, namely, log SSFR > -11 yr$^{-1}$ (star-forming; blue) and log SSFR <  -11 yr$^{-1}$ (quiescent; red). Each bin contains $\sim100$ galaxies. The error bars are given by $\sigma/\sqrt{N}$, where $\sigma$ is the standard deviation of $L_{\rm sbar}$ per bin, and $N$ is the total number of galaxies per bin.
\label{fig:barlength_vs_appendix}}
\end{figure*}

\begin{figure*}[t!]  
\centering
\includegraphics[scale=.80]{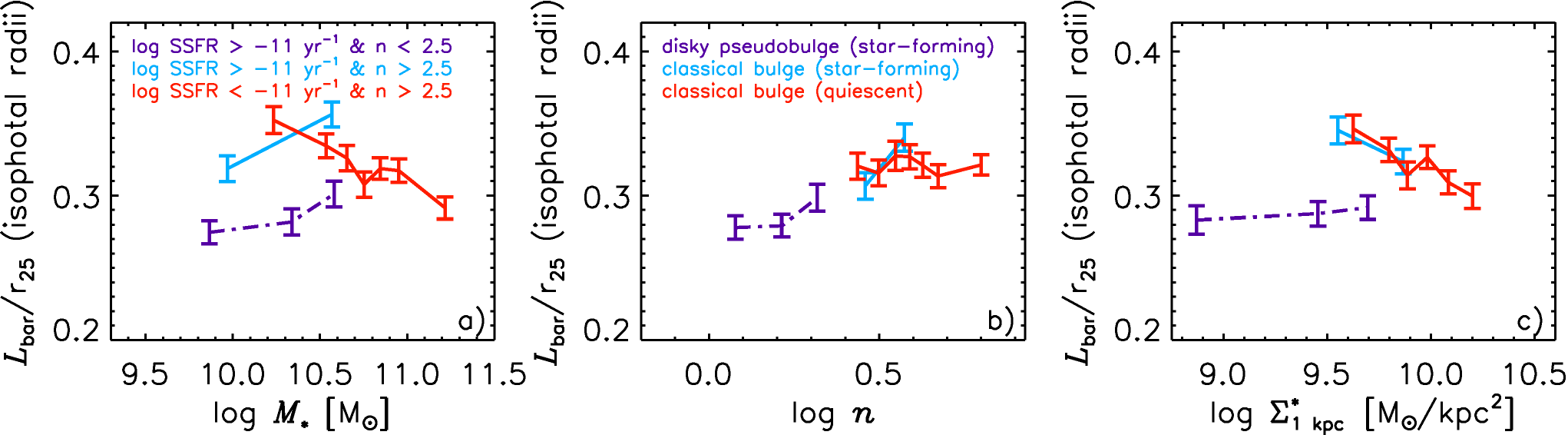}
\caption{Average $L_{\rm bar}/\rm{r_{25}}$ plotted against: {\it a)} $M_*$, {\it b)} $n$, and {\it c)} $\Sigma_{\rm 1~kpc}^*$. The details of this figure are identical to that of Fig.~\ref{fig:barlength_vs_appendix}, with the exception that each bin contains $\sim75$ galaxies and also, galaxies are further separated by bulge type, as identified by $n$. Purple points represent the star-forming disky pseudobulge galaxies, light blue points represent the star-forming classical bulge galaxies, and red points represent the quiescent classical bulge galaxies.
\label{fig:barlength_1d_appendix}}
\end{figure*}

Comparing the trends of bar length scaled by the isophotal radii (Fig.~\ref{fig:barlength_vs_appendix} and \ref{fig:barlength_1d_appendix}) to those of $L_{\rm sbar}$ (Fig.~\ref{fig:barlength_vs} and \ref{fig:barlength_1d}) shows a good agreement. The only noticeable differences are at the highest $n$ and $\Sigma_{\rm1~kpc}^*$, which is hard to interpret and may be due to a number of issues. These GIM2D disk scale lengths may be affected by the prominent bars present in these galaxies. But the better sky background determination and better object deblending of the GIM2D decompositions could also lead to a more accurate measurement of the disk scale length. More work needs to be done to truly understand the differences, but this is outside the scope of this paper. Moreover, this paper does not put a strong emphasis on the bar length trends at the highest $n$ and $\Sigma_{\rm1~kpc}^*$, thus it does not affect our major conclusions.

\section{R90/R50} \label{appendix:r90r50}

\begin{figure*}[t!]  
\centering
\includegraphics[scale=.80]{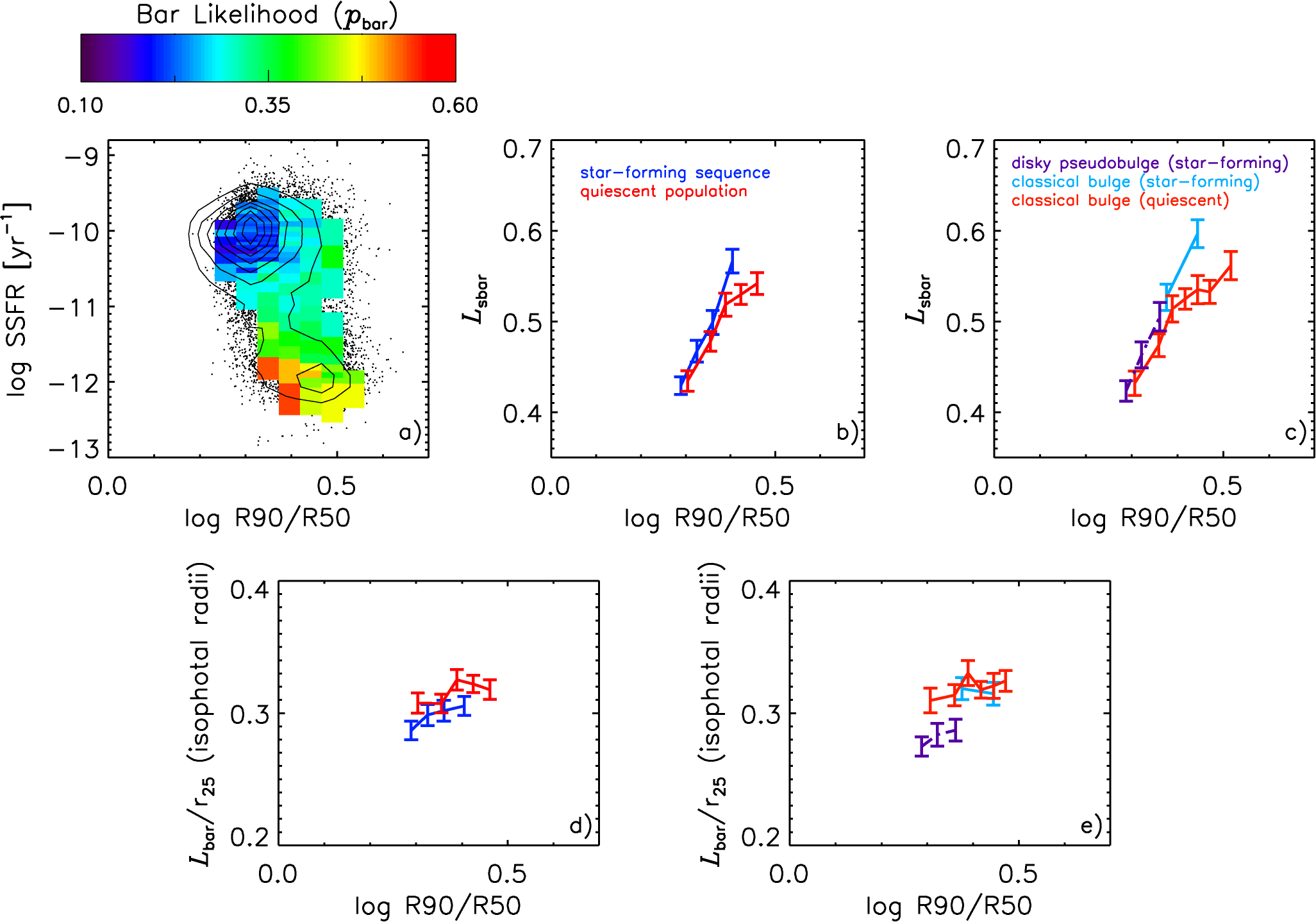}
\caption{Replacing S\'ersic index with R90/R50. {\it a)} plots average $p_{\rm bar}$ in bins of SSFR and R90/R50. {\it b) \& c)} plot bar length scaled by GIM2D bulge+disk model disk scale length vs. R90/R50. {\it d) \& e)} plot bar length scaled by the isophotal radii vs. R90/R50. 
\label{fig:r90r50}}
\end{figure*}

Fig.~\ref{fig:r90r50} shows the effects of using the Petrosian concentration index from SDSS, R90/R50, where R90 and R50 are the radii enclosing 90 and 50 \% of the galaxy luminosity, respectively. The trends with $p_{\rm bar}$ in Fig.~\ref{fig:r90r50}a are almost identical to that with S\'ersic index (Fig.~\ref{fig:barfrac}b). Fig.~\ref{fig:r90r50}b and \ref{fig:r90r50}c show that the trends with R90/R50 for the star-forming sequence, star-forming disky pseudobulge galaxies, and the star-forming classical bulge galaxies are the same as with $n$ (Fig.~\ref{fig:barlength_vs}b and \ref{fig:barlength_1d}b), i.e., $L_{\rm sbar}$ increases with increasing $n$ or R90/R50. 

For the quiescent population (Fig.~\ref{fig:r90r50}b) and the quiescent classical bulge galaxies (Fig.~\ref{fig:r90r50}c), however, there is a noticeable difference between the trends of $L_{\rm sbar}$ at the highest values of R90/R50 and $n$. Namely, while there is a decrease of $L_{\rm sbar}$ at the highest $n$ (Fig.~\ref{fig:barlength_vs}b and \ref{fig:barlength_1d}b), there seems to be a steady increase of $L_{\rm sbar}$ with increasing R90/R50. It is unclear why this is the case. It could be due to the improved sky background determination and object deblending in the fits of \cite{simard11} compared to the standard SDSS pipeline. However, no matter the reason, this minor difference does not affect the paper since we leave the interpretation of the bar length trends for the highest $n$ values open.

Comparing the trends of bar length scaled by the isophotal radii between $n$ (Fig.~\ref{fig:barlength_vs_appendix}b and \ref{fig:barlength_1d_appendix}b) and R90/R50 (Fig.~\ref{fig:r90r50}d and \ref{fig:r90r50}e) shows general agreement between all populations.

Thus the results from R90/R50 and $n$ are largely similar, and the use of either would not change the main conclusions of the paper.

%% ------------------------------------------------------------------
%% REFERENCES
%% ------------------------------------------------------------------

%\bibliography{mips} \bibliographystyle{apj}

\end{document}